%% file: main.tex
\documentclass[sigplan,screen,nonacm]{acmart}

\renewcommand\footnotetextcopyrightpermission[1]{}
\microtypesetup{expansion=false}

\usepackage{amsmath}

\usepackage{amssymb}
\usepackage{pifont}
\usepackage{booktabs}
\usepackage{multirow}
\usepackage{colortbl}
\usepackage{graphicx}
\usepackage{xcolor}
\usepackage{subcaption}
\usepackage{placeins}
\usepackage{threeparttable}

\definecolor{headerblue}{HTML}{E3F2FD}
\definecolor{lightgreen}{HTML}{E8F5E9}

\title{ObjectCache: Layerwise Object-Storage Retrieval for KV Cache Reuse}

\author{Yu Zhu}
\affiliation{%
  \institution{ETH Zurich}
  \country{Switzerland}
}
\email{yu.zhu@inf.ethz.ch}

\author{Aditya Dhakal}
\affiliation{%
  \institution{HPE Labs}
  \country{United States of America}
}
\email{aditya.dhakal@hpe.com}

\author{Yunming Xiao}
\affiliation{%
  \institution{The Chinese University of Hong Kong, Shenzhen}
  \country{China}
}
\email{yunmingxiao@cuhk.edu.cn}

\author{Dejan Milojicic}
\affiliation{%
  \institution{HPE Labs}
  \country{United States of America}
}
\email{dejan.milojicic@hpe.com}

\author{Gustavo Alonso}
\affiliation{%
  \institution{ETH Zurich}
  \country{Switzerland}
}
\email{alonso@inf.ethz.ch}

\begin{document}

\begin{abstract}
\input{sections/00_abstract}
\end{abstract}

\maketitle

\input{sections/01_introduction}

\input{sections/02_background}

\input{sections/03_design}
\input{sections/04_implementation}
\input{sections/05_evaluation}
\input{sections/06_discussion}
\input{sections/07_related_work}

\input{sections/08_conclusion}

\bibliographystyle{ACM-Reference-Format}
\bibliography{references}

\clearpage
\input{sections/09_appendix}

\FloatBarrier
\end{document}

%% file: sections/00_abstract.tex
Prefix KV caching has become a key mechanism in LLM serving: it reduces time to first token (TTFT) by avoiding redundant computation across requests that share a prefix (i.e., the system prompt). 
However, the accumulated KV cache is often larger than what GPU memory and local DRAM can hold. 
To preserve latency, current systems keep the KV cache in remote DRAM pools, increasing serving-cluster size and cost. 
In this paper, we explore a different approach: storing the KV cache in S3-compatible object storage so that capacity is no longer the constraint, while minimizing the impact on TTFT. 
We propose ObjectCache, which co-designs the storage protocol and transfer schedule so that the storage server delivers KV cache data in the order the GPU consumes it, overlapping data transfer with compute across concurrent requests.
We prototype ObjectCache on a 100 Gbps RoCE cluster with NIXL (an inference library that abstracts storage and memory), Ceph RGW (an Object Gateway for clusters), and DAOS (an open source storage system). 
For 64K contexts, common in today's systems, ObjectCache adds only 5.6\% latency over local DRAM; for 4K contexts, where less compute is available to mask transfer, ObjectCache adds 56--75\,ms over the optimal local layerwise baseline. 
Under shared bandwidth caps, our scheduler reduces added TTFT by 1.2--1.8x compared with equal bandwidth sharing.

%% file: sections/01_introduction.tex
\section{Introduction}
\label{sec:intro}
\begin{figure}[t]
    \centering
    \includegraphics[width=0.95\columnwidth,pagebox=cropbox]{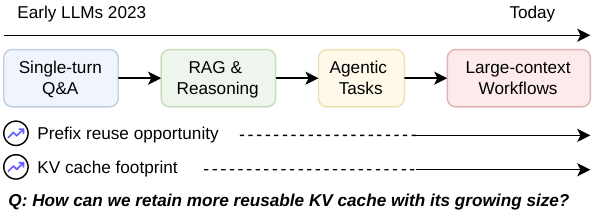}
    \caption{Long-context LLM tasks increasingly reuse long-lived prefixes,
    growing the aggregate KV cache footprint that a serving cluster must retain.}
    \label{fig:llm_task_evolution}
\end{figure}

Modern LLM serving increasingly relies on long and reusable input contexts, including system prompts, retrieval-augmented generation, code repositories, long conversations, and agent histories \cite{xiang2026servegen, wang2025burstgpt, fan2024survey, abou2025agentic} (Figure~\ref{fig:llm_task_evolution}).
These workloads make the prefill phase expensive because the model must materialize key--value (KV) tensors for every token and every layer before decoding can begin.
Prefix KV-cache reuse avoids recomputing the shared prefix across requests and is therefore a primary mechanism for reducing TTFT.
As context lengths and reuse opportunities grow, however, the aggregate KV-cache footprint that a serving cluster must retain also grows rapidly (Appendix Figure~\ref{fig:repo_kv_range}).

The challenge is that reusable KV cache is much larger than the memory capacity naturally available near GPUs.
GPU memory is scarce and needed for model weights and active requests, while local DRAM only scales with the serving nodes that happen to produce or consume the cache.
Recent systems therefore use remote DRAM pools or memory-centric KV-cache tiers to preserve low latency \cite{qin2024mooncake, hu2024memserve}.
Although effective, this approach makes cache capacity a provisioned part of the serving cluster: memory must be sized for the retained KV working set and remains coupled to the compute deployment, thus increasing cost and limiting how much long-lived reusable context the system can retain.

This motivates a different question: can S3-compatible object storage serve as a runtime backend for reusable KV cache?
Object storage is attractive because prefix KV blocks are immutable after prefill, naturally addressable by content-derived prefix hashes, and reusable across users, sessions, and compute nodes.
Unlike a remote-DRAM tier, an object store scales capacity independently of the GPU serving cluster and persists cache state beyond the lifetime of any particular worker.
If object storage could be used on the serving path, prefill and decode workers would become largely stateless with respect to reusable prefixes: once KV blocks are committed, later requests could fetch them from the shared object tier rather than returning to the node that produced them \cite{zhong2024distserve,patel2024splitwise}.

The obstacle is not only transport bandwidth.
Even with an RDMA data path like Nvidia cuObject \cite{cuobject, cuobject_server}, the standard S3 abstraction remains mismatched to runtime KV-cache reuse.
S3 exposes object-level operations: a request names one object, or one byte range of one object.
In contrast, a prefix hit returns a set of many hash-addressed KV chunks, and the inference engine consumes the matched cache in layer order.
Fetching each chunk or each layer range with independent S3 requests exposes fixed request overhead on the TTFT critical path, while fetching coarse objects sacrifices fine-grained prefix reuse.
Thus, the gap between S3 and runtime KV-cache reuse is semantic: object storage moves objects, whereas inference needs prefix-selected KV blocks delivered in the order the GPU consumes them.

We present ObjectCache, a protocol-scheduling co-design that makes S3-compatible object storage suitable for runtime KV-cache reuse.
On the protocol side, ObjectCache keeps KV cache stored as fine-grained, hash-addressed chunks, but extends the S3-compatible request with a compact descriptor that names the matched chunks, model layout, delivery order, and RDMA target.
The storage server uses this descriptor to gather many chunk ranges, assemble one layer-major payload at a time, and deliver each layer directly to the serving node over RDMA.
On the scheduling side, ObjectCache exploits the fact that layerwise transfer can overlap with per-layer GPU computation.
Rather than sharing bandwidth equally or in proportion to bytes, ObjectCache allocates bandwidth according to each request's per-layer stall target, avoiding bandwidth assignments that do not further reduce TTFT.

We prototype ObjectCache on a 100~Gbps RoCE cluster using NIXL \cite{nixl}, Ceph RGW \cite{ceph}, and DAOS \cite{daos}, and evaluate it with Llama~3.1~8B \cite{llama_3.1_8B}.
For 64K-token contexts, ObjectCache adds only 5.6\% TTFT over an optimized local layerwise baseline.
For shorter 4K-token contexts, where less compute is available to hide transfer, ObjectCache adds 56--75~ms over the local layerwise baseline.
Under shared bandwidth caps, the ObjectCache scheduler reduces added TTFT by 1.2--1.8$\times$ compared with equal bandwidth sharing.

This paper makes the following contributions:
\begin{itemize}
    \item We propose ObjectCache, a protocol-scheduling co-design for serving reusable KV cache from S3-compatible object storage.
    \item We design an S3-compatible protocol extension that supports prefix-selected multi-object aggregation and layerwise KV delivery while preserving fine-grained client-side prefix lookup.
    \item We introduce a bandwidth scheduler that allocates storage bandwidth according to per-layer compute-transfer overlap, reducing TTFT under contention.
    \item We demonstrate on a 100~Gbps RoCE prototype that ObjectCache approaches local layerwise KV-cache performance for long-context workloads and improves TTFT under shared bandwidth limits.
\end{itemize}

%% file: sections/02_background.tex
\section{Background and Motivation}
\label{sec:background}

\begin{figure}[t]
    \centering
    \includegraphics[width=0.95\columnwidth]{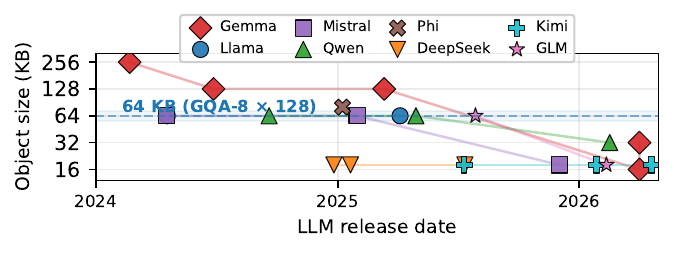}
    \caption{Per-layer KV payload for a 16-token chunk across recent open-weight
    LLM families. The 64 KB dashed line marks the grouped-query attention (GQA)
    baseline with 8 KV heads of 128 dimensions; multi-head latent attention
    (MLA) and smaller head counts push recent models below this threshold.}
    \label{fig:model_kv_timeline}
\end{figure}

\subsection{KV cache and Prefix Reuse}

During the prefill phase, LLM inference processes the input prompts and materializes the per-layer
key/value tensors to be used in transformer attention
\cite{vaswani2017attention,dao2022flashattention}. During the decode phase, the model reuses these tensors instead of recomputing attention over the entire prefix each time by storing the context in a KV cache.
To enable cross-query reuse and storage, serving systems partition 
the KV cache into fixed-size chunks corresponding to $G$ consecutive tokens.
For a model with $L$ layers, $n_{\mathrm{kv}}$ KV heads, head dimension $d$, and
element width $p$, the KV bytes per token and the per-layer bytes of a
$G$-token chunk are:
\begin{equation}
\label{eq:kv-bytes}
    \mathrm{KV}_{\mathrm{token}} =
    2L n_{\mathrm{kv}} d p,
    \qquad
    S_{\mathrm{layer,chunk}} =
    2G n_{\mathrm{kv}} d p .
\end{equation}
The prefix KV cache reuse relies on a radix tree or similar
prefix index to find the longest cached match for a new request
\cite{kwon2023efficient,ye2024chunkattention,gim2024prompt,gao2024cost,wang2025kvcache}.
Each chunk can be identified by a rolling hash,
$H_i = \mathrm{Hash}(H_{i-1} \parallel \mathrm{tokens}_i)$, that gives it a
deterministic object key. Using $H_i$ as the object key gives KV chunks the
properties object storage is good at: immutable writes, content-addressed
deduplication, and independent reuse across requests.

\begin{figure}[t]
\centering
\begin{subfigure}[t]{0.43\columnwidth}
    \centering
    \includegraphics[width=\linewidth]{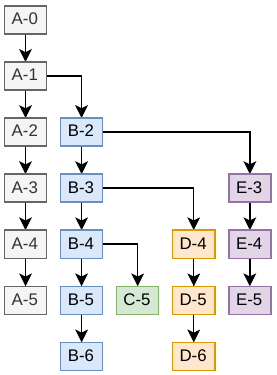}
    \caption{Fine granularity preserves intermediate branch points.}
    \label{fig:radix_fine}
\end{subfigure}\hfill
\begin{subfigure}[t]{0.50\columnwidth}
    \centering
    \includegraphics[width=\linewidth]{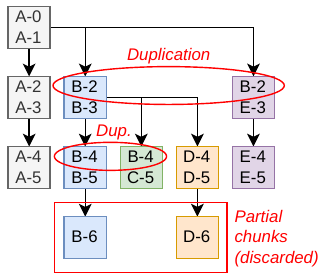}
    \caption{Coarse granularity merges branch points.}
    \label{fig:radix_coarse}
\end{subfigure}
\caption{Prefix reuse under fine and coarse storage granularities. Coarse chunks
reduce index depth but lose branch points where requests can diverge, forcing
otherwise reusable tokens to be recomputed.}
\label{fig:radix_tree_granularity}
\end{figure}

\begin{figure}[t]
\centering
\includegraphics[width=0.95\columnwidth]{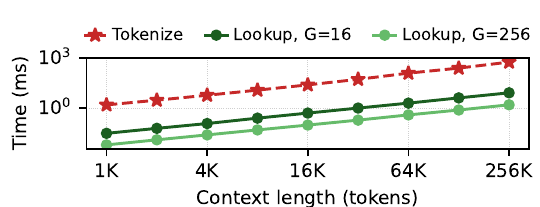}
\caption{Prefix-hash lookup cost is small relative to tokenization, even at
16-token granularity. Fine-grained indexing is therefore not the request
critical-path bottleneck.}
\label{fig:hash_lookup}
\end{figure}

\begin{figure*}[t]
\centering
\includegraphics[width=0.95\textwidth]{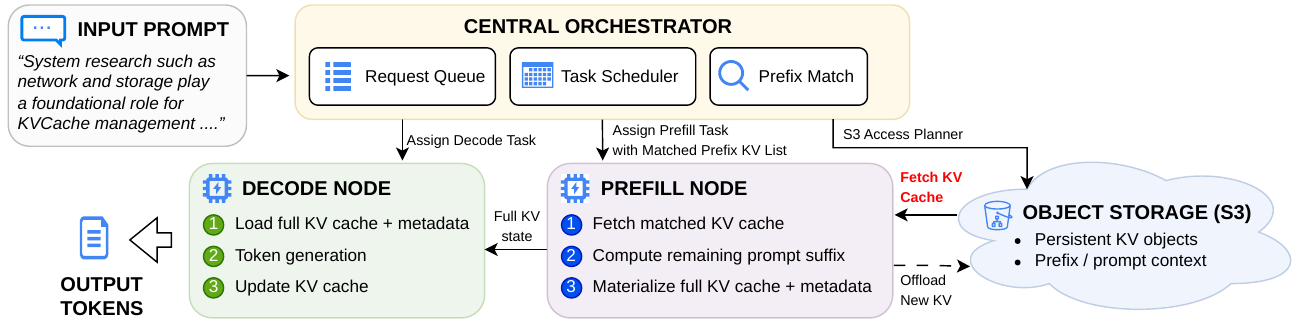}
\caption{ObjectCache in a disaggregated cluster. Prefix KV caches are stored in an
object-storage tier through an S3-compatible interface, decoupling prefill and
decode workers from the machines that produced the cache. ObjectCache extends this
interface so object storage can serve KV cache reuse with the granularity and
layerwise delivery order expected by LLM serving systems.}
\label{fig:ObjectCache_motivation}
\end{figure*}

At fine chunk sizes, the per-layer KV payload becomes small enough that the fixed
object-request overhead is visible even when the communication overhead is small
(Figure~\ref{fig:model_kv_timeline} instantiates
Equation~\ref{eq:kv-bytes} for current models). A 16-token chunk exposes only
tens of KB per layer for many current models, and MLA-style layouts can reduce
the layer slice to roughly 16~KB \cite{liu2024deepseek, meng2026transmla}.
This pressure remains even at coarser chunk sizes. For Llama 3.1 8B \cite{llama_3.1_8B}, a
256-token chunk is about 1~MB per layer; compact KV layouts, MLA-style
representations, or shape-preserving compression can push that below the
efficient object-transfer regime \cite{chang2025palu, zandieh2026turboquant}. Increasing the chunk granularity enables larger physical data
transfers, but also coarsens prefix reuse. ObjectCache instead keeps the logical
reuse granularity independent of the effective transfer granularity by
aggregating chunks at the storage server.

Figures~\ref{fig:radix_tree_granularity} and~\ref{fig:hash_lookup} separate two
concerns that are often conflated. Figure~\ref{fig:radix_tree_granularity}
shows why fine chunks preserve more radix-tree branch points and avoid
unnecessary recompute after a request diverges. 
Figure~\ref{fig:hash_lookup} shows that prefix-hash lookup 
remains small compared to tokenization, even at 
$G{=}16$ granularity. Fine-grained indexing does not 
add meaningful overhead to the request critical path.
The system bottleneck is therefore not
finding the prefix; it is delivering the matched KV chunks in the order the
model can consume.

\subsection{Serving-Path KV Access Pattern}
In disaggregated serving, a prefill worker materializes KV cache for the input prompt, while a decode worker later uses the resulting KV state to generate tokens.
When a new request shares a cached prefix, prefix lookup returns an ordered list of matched KV chunks that the prefill worker can reuse before computing the remaining suffix.
Although these chunks are stored independently, the model consumes them layer by layer: it needs the layer~0 slice from all matched chunks before computing layer~0, then the layer~1 slice before computing layer~1, and so on.
Thus, the storage-facing access pattern is an ordered multi-object, multi-range read rather than a single object read.

%% file: sections/03_design.tex
\section{ObjectCache Design}
\label{sec:design}


Figure~\ref{fig:ObjectCache_motivation} shows how ObjectCache fits into a disaggregated LLM serving cluster.
A central orchestrator receives a request, performs prefix matching, and assigns the remaining prefill work to a prefill node together with the matched prefix KV list.
The prefill node fetches the reusable KV cache from the S3-compatible object tier, computes the remaining prompt suffix, and materializes the full KV state.
The decode node can then load the full KV state and generate output tokens, while newly produced KV blocks are offloaded back to object storage for future reuse.
This organization decouples prefill and decode workers from the machines that originally produced the cache, making reusable prefix state persistent and shared through the object-storage tier.

The key challenge is that this serving path needs more than ordinary object reads.
ObjectCache is built around one design rule: \emph{keep storage objects small enough for prefix reuse, but expose a serving interface that returns large, layer-ordered transfers}.
The design is therefore an interface extension, not a new transport.
The underlying S3-over-RDMA data path moves bytes; ObjectCache changes what the serving system can ask the object tier to do.


\begin{figure}[t]
\centering
\includegraphics[width=0.95\columnwidth]{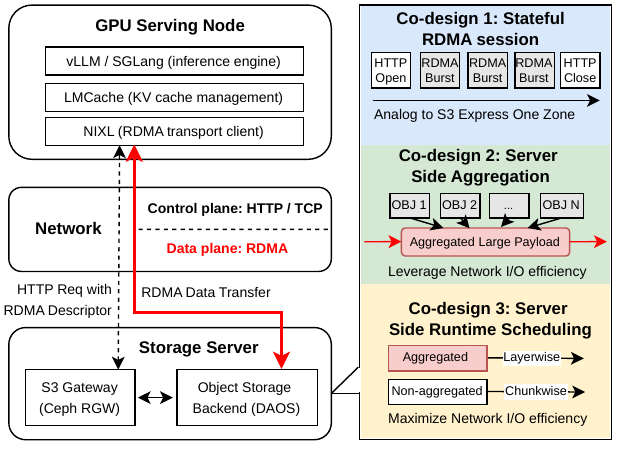}
\caption{ObjectCache system design. HTTP preserves S3-compatible control, while the
storage server aggregates matched chunk ranges into layerwise KV payloads and
delivers them to the serving node over RDMA.}
\label{fig:architecture}
\end{figure}


The resulting architecture has three roles (Figure~\ref{fig:architecture}).
The \emph{GPU serving node} issues S3-compatible requests and consumes KV cache data
through the inference engine's layerwise interface.
The \emph{gateway} terminates the S3 control plane, parses the ObjectCache
descriptor, forwards the multi-object request to the storage server, and
upgrades the S3 control plane to express prefix-aware, layer-scheduled KV
transfer.
The \emph{storage server} resolves chunk objects in the S3 namespace,
performs range reads, assembles layer-major payloads, RDMA-writes them
directly to the client buffer, and executes the layer-scheduled semantic at
the object/range level.
All runtime policy---whether a given request is served chunkwise or via
layerwise aggregation, and how concurrent tenants share the network
cap---runs on the storage server, so the gateway and NIXL client both remain
stateless with respect to scheduling decisions
(Sections~\ref{sec:design:mode-selection}
and~\ref{sec:design:scheduling}).

\begin{table}[t]
\caption{ObjectCache request descriptor.}
\label{tab:ObjectCache_descriptor}
\centering
\footnotesize
\setlength{\tabcolsep}{4pt}
\renewcommand{\arraystretch}{1.08}
\begin{tabular}{@{}p{0.4\columnwidth}p{0.55\columnwidth}@{}}
\toprule
\textbf{Field} & \textbf{Meaning} \\
\midrule
\texttt{chunk\_keys} & Matched prefix chunks
$[H_0,\ldots,H_{N-1}]$. \\
\texttt{num\_layers} & $L$: Number of model layers. \\
\texttt{chunk\_tokens} & $G$: Tokens per stored chunk. \\
\texttt{per\_layer\_chunk\_bytes} & $S$: Slice size inside each chunk. \\
\texttt{delivery} & Layer-major order. \\
\texttt{rdma\_target} & Client buffer address, key, and length. \\
\bottomrule
\end{tabular}
\end{table}

\subsection{Required Semantics}
\label{sec:design:semantics}

ObjectCache requires four access semantics that no existing S3 primitive
provides in combined form; this is the gap between accelerating S3 through better transport protocols and
making S3 suitable for prefix KV reuse.

\textbf{(1) Multi-object batching.}
A prefix hit returns a list of matched KV chunks, not one object. If each chunk
requires an independent S3 request, fixed request overhead dominates in
the fine-grained regime that prefix caching prefers.

\textbf{(2) Layerwise delivery.}
Inference consumes KV cache data layer by layer. Delivering complete chunks in chunk-major
order forces the NIXL client to wait for the full matched prefix before starting
layer 0. Delivery of layer 0 for all chunks first creates a compute window to transfer later layers.

\textbf{(3) Hash-derived immutable keys.}
KV chunks must remain addressable by hashes so that 
requests reuse the same object when they share a prefix. This preserves the
radix-tree storage semantics used by existing prefix caches.

\textbf{(4) Bulk-transfer compatibility.}
Once the layer-major payload has been assembled, data should be transferred using
RDMA because it provides the high-throughput transport that the higher-level aggregation
semantics can use.

Standard S3 GET satisfies hash-derived keys but not batching or layerwise
delivery. Range-GET can fetch one layer range, but only by issuing a request
per chunk per layer. S3-over-RDMA satisfies the bulk-transfer requirement, but
it inherits the same single-object request model. ObjectCache adds the missing
server-side multi-object aggregation and layerwise delivery semantics while
reusing the RDMA.

\subsection{The ObjectCache Descriptor}
\label{sec:design:streaming}

ObjectCache adds an S3-compatible descriptor to a normal request. The descriptor names the matched chunk keys, the model layout, the delivery order, and the RDMA target buffer.

Table~\ref{tab:ObjectCache_descriptor} lists the fields of the descriptor, which is intentionally arithmetic rather than manifest-heavy.
Because every chunk in the same model deployment has the same per-layer size
$S$, the byte range for layer $\ell$ in a chunk is $[\ell S,\;(\ell+1)S)$.
Thus, the storage server can derive all fixed-shape layer ranges from the
descriptor without a per-object manifest.
Variable-size or compressed layouts can add a manifest later, but the common
serving fast path keeps the descriptor compact.

\subsection{Server-Side Layer Aggregation}
\label{sec:design:aggregation}

The storage server executes the descriptor by assembling one payload per model
layer. For each layer $\ell \in [0,L)$, it range-fetches
$[\ell S,(\ell+1)S)$ from every matched chunk in parallel, appends the returned
slices in prefix order, RDMA-writes the assembled payload to the client buffer,
and notifies the serving node that the layer is ready (pseudocode in Appendix
Table~\ref{alg:layerwise-get}). This notification is on the critical path of
inference: it lets the GPU start layer computation without waiting for the
whole prefix to arrive \cite{qin2024mooncake, liu2025lmcache}.

The physical storage layout remains chunkwise even though the logical delivery
layout is layerwise. In KV\_L2TD format (\textbf{K}ey-\textbf{V}alue tensors
stored in \textbf{L}ayer-major order, with the \textbf{2} matrices
concatenated per layer, then \textbf{T}oken position, then hidden
\textbf{D}imension), each immutable prefix-chunk object stores all layers
sequentially, and each layer is arranged by token position and hidden
dimension. Server-side aggregation changes only the readout order: one
layerwise payload covers all matched chunks for a single model layer.

Keeping storage fine-grained matters because coarsening sacrifices
reuse. Larger KV blocks reduce object retrieval overhead but discard reuse when two
requests diverge inside a block; increasing the lookup granularity from 16
to 512 tokens can force otherwise reusable tokens to be recomputed across
models and hit boundaries (Appendix
Table~\ref{tab:granularity_comparison} quantifies this recompute cost).
The right design point is therefore not ``make objects as large as
possible''; the system should keep lookup granularity small and transfer granularity large enough for
efficient I/O.

\subsection{Server-Side Mode Selection}
\label{sec:design:mode-selection}

ObjectCache does not serve every cache hit using server-side layer aggregation.
The KV cache is always stored in the chunkwise layout above, so the storage
server chooses only the delivery mode. Each incoming descriptor names the
matched chunks, from which the server derives the total matched payload
$W = N \cdot L \cdot S$, where $N$ is the number of matched chunks, $L$ is the
number of model layers, and $S$ is the per-layer chunk size. It then applies a
single size-threshold rule:
\begin{equation}
\label{eq:mode-dispatch}
    \mathrm{mode}(W) =
    \begin{cases}
        \text{chunkwise} & W < \Theta, \\
        \text{layerwise + aggregation} & W \ge \Theta.
    \end{cases}
\end{equation}
The threshold $\Theta$ is a deployment knob: payloads below $\Theta$ use
chunkwise delivery, and larger payloads use server-side aggregation with
layerwise delivery. Below~$\Theta$, the layerwise schedule of
Equation~\ref{eq:layerwise-ttft} still minimizes TTFT in theory, but the
absolute saving is negligible relative to the aggregation overhead.
Above~$\Theta$, the aggregation pipeline recovers its CPU and I/O-depth cost as
measurable TTFT reduction while also amortizing per-object descriptor overhead
across many small range reads.

As an illustration, on our 100~Gbps prototype with Llama 3.1 8B we use
$\Theta \approx 512$~MB, the payload size at which network transfer time at
line rate becomes comparable to the prefill compute window. Under this rule,
the 4K configurations in Section~\ref{sec:eval} fall on the chunkwise side and
the 16K/64K configurations fall on the aggregated-layerwise side, matching
where each mode wins in Figure~\ref{fig:ttft_delta_over_best}. The appropriate
value of $\Theta$ depends on the storage backend, SSD queue depth, and link
rate; the dispatch rule itself does not. A full sensitivity analysis of
$\Theta$ is left to future work.

Equation~\ref{eq:mode-dispatch} is also what decides which requests enter
multi-tenant scheduling: chunkwise requests run independently against the
object storage and consume only their fair share of object-store concurrency,
while every layerwise request joins the shared bandwidth pool described
next. The scheduling problem below is therefore scoped to layerwise requests
under a shared bandwidth cap.

\subsection{Layerwise Prefetch}
\label{sec:design:prefetch}


\begin{figure}[t]
\centering
\begin{subfigure}[t]{\columnwidth}
  \centering
  \includegraphics[width=\columnwidth]{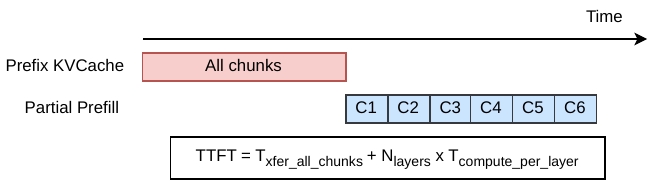}
  \caption{Chunkwise loading waits for the full cached prefix before compute can start.}
  \label{fig:scheduler:chunkwise}
\end{subfigure}\\[0.3em]
\begin{subfigure}[t]{\columnwidth}
  \centering
  \includegraphics[width=\columnwidth]{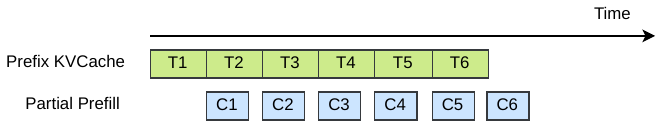}
  \caption{Transfer-bound layerwise prefetch.}
  \label{fig:scheduler:prefetch}
\end{subfigure}\\[0.3em]
\begin{subfigure}[t]{\columnwidth}
  \centering
  \includegraphics[width=\columnwidth]{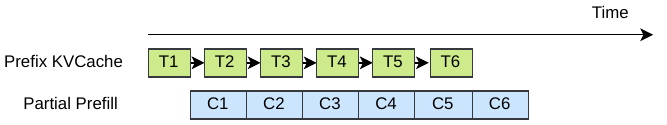}
  \caption{Compute-bound layerwise prefetch.}
  \label{fig:scheduler:naive}
\end{subfigure}
\caption{KV cache fetch scheduling. Chunkwise delivery serializes cache loading
before prefill, while layerwise delivery exposes per-layer readiness so transfer
can overlap compute when bandwidth is sufficient and otherwise appears as
per-layer stall.}
\label{fig:ObjectCache_scheduler}
\end{figure}

Figure~\ref{fig:ObjectCache_scheduler} explains why the delivery order matters.
A chunkwise baseline waits for all matched chunks before compute can use
any layer from the KV cache.  Layerwise prefetch instead transfers layer~0
first; once layer~0 arrives, the GPU begins computing on it while the
remaining layers transfer in parallel.  Let $X_\ell$ be the transfer time
for layer~$\ell$ and $C_\ell$ be the compute time exposed by the miss
tokens at layer~$\ell$.  With one-layer prefetch, the TTFT model is
\begin{equation}
\label{eq:layerwise-ttft}
    T_{\mathrm{TTFT}}
    \approx X_0
    + \sum_{\ell=0}^{L-2} \max(X_{\ell+1},\, C_\ell)
    + C_{L-1},
\end{equation}
where $X_0$ is the latency before the GPU can start (layer~0 must
arrive in full), the summation captures $L{-}1$ stages in which
transfer and compute overlap, and $C_{L-1}$ is the final layer's
compute after all transfers have finished.
When $X_\ell > C_\ell$, the transfer time that exceeds the compute
window appears as additional waiting time in the corresponding stage
of Equation~\ref{eq:layerwise-ttft}.
Section~\ref{sec:eval:overlap} evaluates this model with measured
compute times and ObjectCache throughput.

\subsection{Bandwidth-Aware Scheduling}
\label{sec:design:scheduling}

Under the layerwise delivery of Section~\ref{sec:design:prefetch}, each
request's per-layer transfer can overlap with its per-layer compute window.
This creates a bandwidth-scheduling problem specific to KV cache reuse: under
a shared bandwidth cap~$B$, policies proportional to matched bytes or
equal-share misallocate. Matched-bytes allocation over-serves long-prefix
requests whose per-layer transfer is already shorter than per-layer compute,
so extra bandwidth yields no further benefit, while equal-share under-serves
short-prefix requests whose compute window is small.

The key observation is that each layer of a request transfers roughly the
same KV bytes and performs similar compute, so we characterise request~$i$
by its per-layer transfer size~$s_i$ and per-layer compute
window~$c_i$.\footnote{Both are approximately constant across layers
because every layer has the same number of KV heads and the same
attention/feed-forward network (FFN) structure.}
When the scheduler assigns bandwidth~$r_i$ to request~$i$, the
per-layer transfer time is $s_i/r_i$. Transfer that exceeds the compute
window introduces additional stall:
\begin{equation}
\label{eq:per-layer-stall}
    \tau_i(r_i) \;=\;
    \max\!\Bigl(0,\;\frac{s_i}{r_i} - c_i\Bigr).
\end{equation}
%


\begin{table*}[t]
\centering
\caption{Interface-level positioning. Existing systems accelerate either the
memory tier or the S3 byte path; ObjectCache adds KV-aware multi-object,
layer-major aggregation inside an S3-compatible request.}
\label{tab:sge_vs_aggregation}
\renewcommand{\arraystretch}{1.12}
\setlength{\tabcolsep}{4.5pt}
\small
\begin{tabular}{@{}l l c c c c@{}}
\toprule
\rowcolor{headerblue}
\textbf{Interface} & \textbf{Primary abstraction} &
\textbf{S3 Namespace} & \textbf{Multi-object} &
\textbf{Layer-major} & \textbf{Rate target} \\
\midrule
Remote KV transfer~\cite{qin2024mooncake,hu2024memserve}
& Memory-resident KV blocks
& No & Runtime-specific & Yes & Runtime-specific \\
S3-over-RDMA~\cite{cuobject,cuobject_server,minio_s3_rdma,nvidia_s3_over_rdma,vast_s3_over_rdma}
& One S3 object or byte range
& Yes & No & No & No \\
RDMA scatter-gather
& Registered memory segments
& No & No & No & No \\
\textbf{ObjectCache}
& Hash-addressed KV chunk objects
& Yes & Yes & Yes & Yes \\
\bottomrule
\end{tabular}
\end{table*}

Each request's additional latency vanishes once $r_i$ reaches its
\emph{zero-stall rate} $r_i^{*} = s_i / c_i$; any bandwidth
beyond this point yields no further latency benefit.
In an unconstrained setting, every request would receive~$r_i^{*}$,
but the total demand $\sum_i r_i^{*}$ may exceed the shared
budget~$B$. The scheduler must then distribute a total cut of
$\sum_i r_i^{*} - B$ across requests. Let $\mathcal{R}$ denote the set of active layerwise requests sharing the storage link, and let $\Delta_i \ge 0$ denote the bandwidth cut applied to request $i$ (so its allocated rate is $r_i = r_i^{*} - \Delta_i$). Allocating these cuts to
minimize total stall leads to the \emph{Stall-opt} problem:

\begin{equation}
\label{eq:stallopt-cut}
    \min_{\{\Delta_i\}}
    \;\sum_{i\in\mathcal{R}} 
    \left(\frac{s_i}{r_i^{*} - \Delta_i} - c_i\right)
    \quad \text{s.t.}\quad
    \sum_{i\in\mathcal{R}} \Delta_i = 
    \sum_{i\in\mathcal{R}} r_i^{*} - B,\;
    0 \le \Delta_i \le r_i^{*}.
\end{equation}


Since $c_i$ does not depend on~$\Delta_i$, substituting 
$r_i = r_i^{*} - \Delta_i$ reduces the objective to 
minimising total transfer time under the bandwidth budget:
\begin{equation}
\label{eq:stallopt-simple}
    \min_{\{r_i\}}
    \;\sum_{i\in\mathcal{R}} 
    \frac{s_i}{r_i}
    \quad \text{s.t.}\quad
    \sum_{i\in\mathcal{R}} r_i = B,\;
    0 < r_i \le r_i^{*}.
\end{equation}

Because $s_i/r_i$ is convex in~$r_i$,
Equation~\ref{eq:stallopt-simple} is a convex program.
When $\sum_i r_i^{*} \le B$, every request receives its
zero-stall rate and all additional TTFT vanishes; when the budget
is tight, the optimal allocation admits a closed-form
solution.
In practice, the analytical $r_i^{*}$
falls on the slope of the TTFT curve rather than at the start of
the flat region (see Section \ref{sec:bandwidth_allocation_for_multi_tenants}). 
\emph{Calibrated Stall-opt} shifts the target by
a small positive offset $\delta$ to ensure the operating point lies on the plateau:
\begin{equation}
\label{eq:calibrated-stallopt}
    \hat{r}_i \;=\; r_i^{*} + \delta,
\end{equation}

Calibrated Stall-opt solves the bandwidth allocation of
Equation~\ref{eq:stallopt-cut} with corrected targets~$\hat{r}_i$ as
upper bounds.  When the total demand~$\sum_i \hat{r}_i$ exceeds
the budget~$B$, the closed-form solution distributes the deficit
to minimize total stall across concurrent requests (pseudocode
in Appendix Table~\ref{alg:calibrated-stallopt}). During each scheduling
epoch, a batch of active layerwise requests is admitted under a fixed total
bandwidth budget, and each request receives a stable target layer-delivery
rate for the duration of its KV load. If one request finishes early, its
released bandwidth returns to the pool for the next scheduling epoch rather
than being redistributed immediately to in-flight requests. This conservative
rule makes per-request transfer times predictable, so the serving node does
not need to react to unexpected bandwidth changes during an epoch.




%% file: sections/04_implementation.tex
\section{Implementation}
\label{sec:impl}

Our prototype integrates ObjectCache into a three-node serving/storage stack. The
serving node runs the LLM and issues KV cache reads. The gateway terminates
S3-compatible requests, parses ObjectCache descriptors, and forwards them to the
storage server, which executes the multi-object aggregation. Concretely, the
prototype uses NIXL as the asynchronous transfer library \cite{nixl}, Ceph RGW as the
S3-compatible gateway \cite{ceph}, and DAOS as the storage server behind the gateway \cite{daos}. The
implementation keeps the gateway thin: the serving node issues normal
S3-compatible requests with additional headers, the gateway handles only S3
control, and the DAOS storage server performs the cross-object range reads
and layer-major assembly.

\subsection{S3-Compatible Paths}
\label{sec:s3_name}
The prototype includes five S3-compatible paths. The first three isolate
transport choices for a single object; the last two exercise the ObjectCache
interface:
\begin{itemize}\itemsep0pt
    \item \textbf{S3TCP}: standard S3 request; object sent via
    HTTP/TCP.
    \item \textbf{S3RDMA Buffer}: S3 request naming one object; the gateway
    stages the payload before the RDMA transfer to the NIXL client.
    \item \textbf{S3RDMA Direct}: S3 request naming one object; the data goes
    through the object-store RDMA path without gateway staging.
    \item \textbf{S3RDMA Batch}: S3 request naming multiple objects; one S3 request carries one HTTP header followed by an RDMA burst, thereby amortizing per-object overhead.
    \item \textbf{S3RDMA Agg}: S3 request naming multiple objects; the storage server assembles layer-major payloads and RDMA-writes them to the NIXL client.
\end{itemize}

We use the following names in the end-to-end TTFT evaluation:
\begin{itemize}
    \item \textbf{Local-DRAM-CW} and \textbf{Local-DRAM-LW}: local CPU-DRAM
          KV cache baselines with chunkwise and layerwise delivery,
          respectively.
    \item \textbf{S3Batch-CW}: the S3-backed chunkwise batched path.
    \item \textbf{S3Agg-LW}: ObjectCache's server-side aggregated layerwise path.
\end{itemize}

\begin{figure*}[t]
    \centering
    \includegraphics[width=0.45\textwidth]{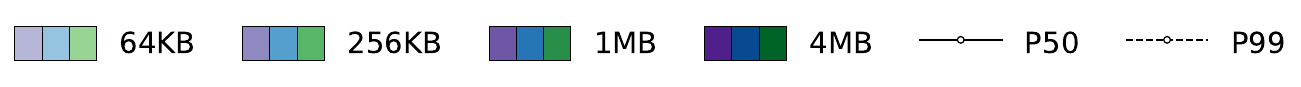}\\[0.3em]
    \begin{subfigure}[t]{0.235\textwidth}
        \centering
        \includegraphics[width=\textwidth]{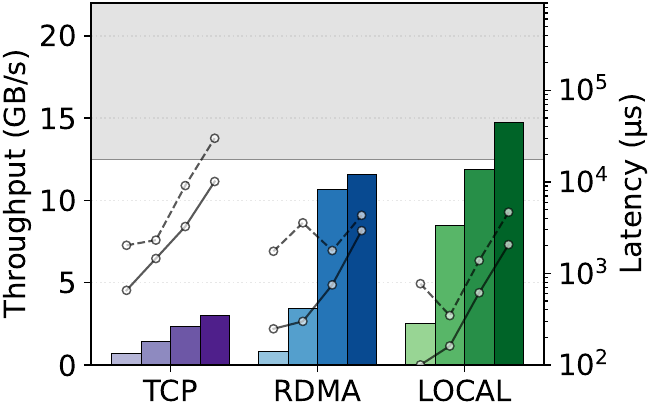}
        \caption{READ, $C{=}8$}
        \label{fig:nixl_dfs_read_1x8}
    \end{subfigure}\hfill
    \begin{subfigure}[t]{0.235\textwidth}
        \centering
        \includegraphics[width=\textwidth]{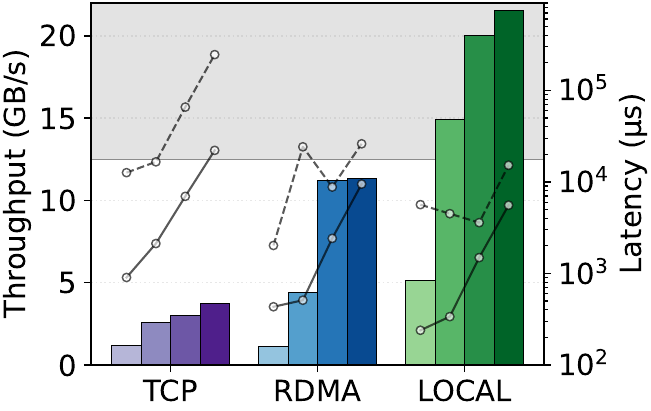}
        \caption{READ, $C{=}32$}
        \label{fig:nixl_dfs_read_4x8}
    \end{subfigure}\hfill
    \begin{subfigure}[t]{0.235\textwidth}
        \centering
        \includegraphics[width=\textwidth]{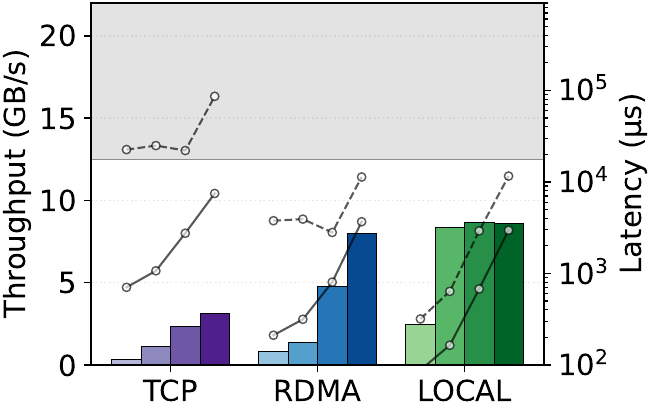}
        \caption{WRITE, $C{=}8$}
        \label{fig:nixl_dfs_write_1x8}
    \end{subfigure}\hfill
    \begin{subfigure}[t]{0.235\textwidth}
        \centering
        \includegraphics[width=\textwidth]{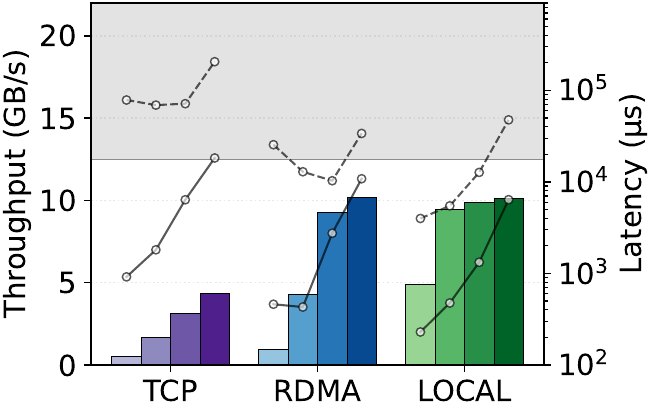}
        \caption{WRITE, $C{=}32$}
        \label{fig:nixl_dfs_write_4x8}
    \end{subfigure}
    \caption{Raw object-storage interface baseline. The gray region marks
    throughput above the 100\,Gbps link capacity; points in this region are
    limited by local storage and host execution rather than the network.}
    \label{fig:nixl_dfs_baseline}
\end{figure*}

\begin{figure*}[t]
    \centering
    \includegraphics[width=0.45\textwidth]{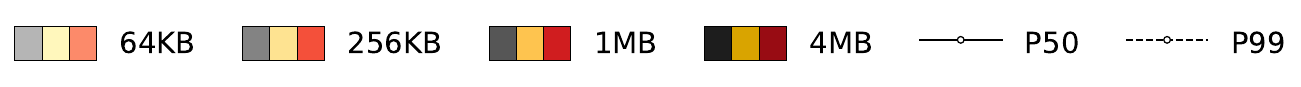}\\[0.3em]
    \begin{subfigure}[t]{0.235\textwidth}
        \centering
        \includegraphics[width=\textwidth]{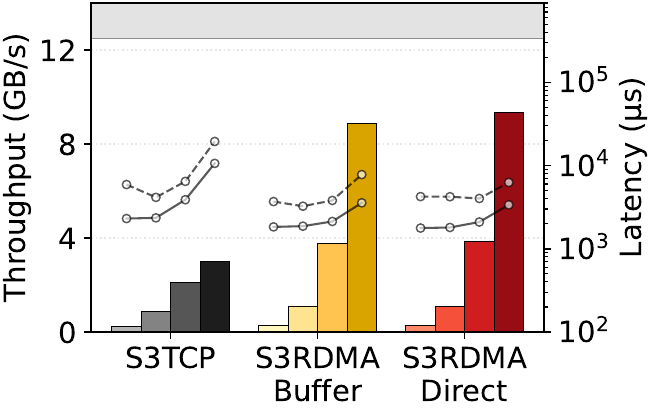}
        \caption{S3 GET, $C{=}8$}
        \label{fig:s3_get_1x8}
    \end{subfigure}\hfill
    \begin{subfigure}[t]{0.235\textwidth}
        \centering
        \includegraphics[width=\textwidth]{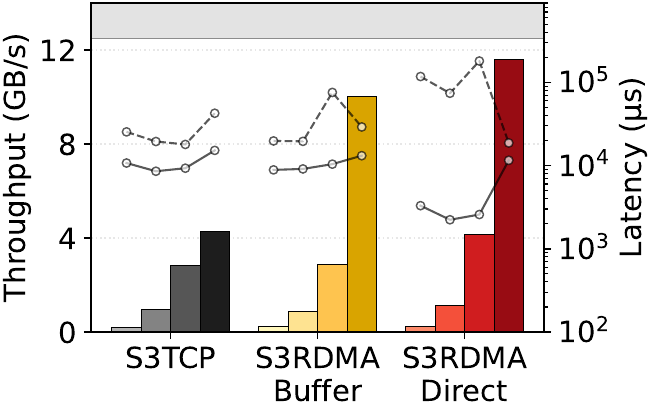}
        \caption{S3 GET, $C{=}32$}
        \label{fig:s3_get_1x32}
    \end{subfigure}\hfill
    \begin{subfigure}[t]{0.235\textwidth}
        \centering
        \includegraphics[width=\textwidth]{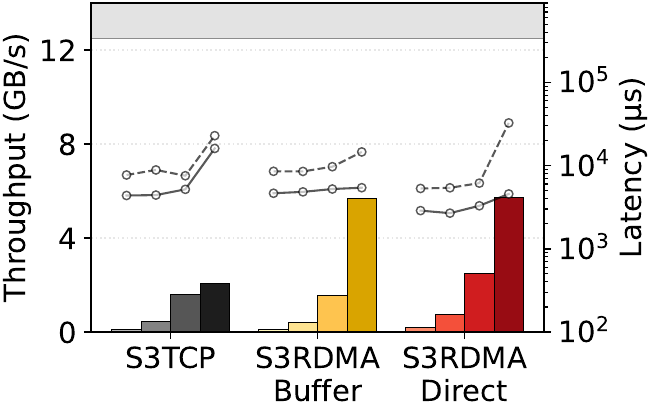}
        \caption{S3 PUT, $C{=}8$}
        \label{fig:s3_put_1x8}
    \end{subfigure}\hfill
    \begin{subfigure}[t]{0.235\textwidth}
        \centering
        \includegraphics[width=\textwidth]{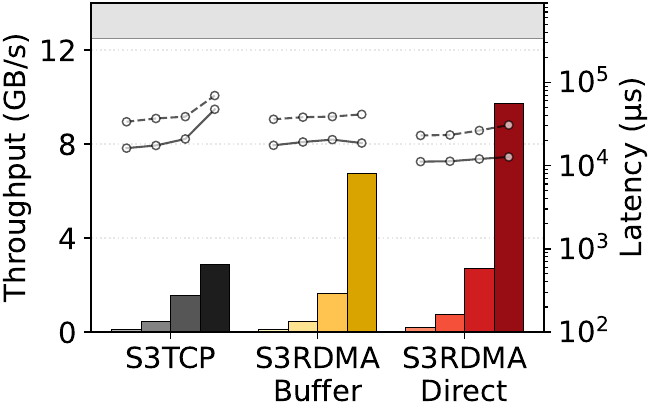}
        \caption{S3 PUT, $C{=}32$}
        \label{fig:s3_put_1x32}
    \end{subfigure}
    \caption{S3-compatible interface baseline. S3RDMA Direct preserves high
    large-object throughput, while S3TCP and S3RDMA Buffer expose protocol and
    staging bottlenecks.}
    \label{fig:s3_throughput}
\end{figure*}

\subsection{GPU Serving Node}

The serving node uses the NIXL object backend to issue S3 operations.
For ordinary objects, the backend can use TCP, an RDMA-buffer mode, or the
RDMA-direct path used in our microbenchmarks. For ObjectCache requests, LMCache \cite{lmcache}
provides the matched prefix chunk keys and the model layout parameters
($L$, $G$, and $S$). NIXL serializes these fields into the
HTTP descriptor and attaches an RDMA target token for the
receive buffer. The LLM inference framework, such as vLLM \cite{vllm},  waits for layer-ready notifications
and lets the model proceed as soon as the next layer's KV has arrived. This is
the same execution pattern used by DRAM-backed layerwise KV loading, but the
source is now object storage.

\subsection{Gateway and Storage Server Execution}

Ceph RGW is used as the gateway. It parses the ObjectCache descriptor after normal S3 request handling, then
forwards the descriptor to the DAOS storage server via a server-side
extension. The DAOS server maps each chunk key to the corresponding DAOS
object, fetches the per-layer record extent from each chunk in
parallel, concatenates the slices in chunk order, and RDMA-writes the
assembled layer payload directly into the client's registered buffer.

The gateway keeps the S3 abstraction in the control path: request headers,
authentication, bucket/object naming, and access control are still handled at
the S3 layer. The data path is split: HTTP carries control, while RDMA carries
the assembled layer payload directly between the storage server and the
GPU serving node. This mirrors the S3-over-RDMA control/data-plane separation
\cite{cuobject,cuobject_server,nvidia_s3_over_rdma}, but ObjectCache adds
multi-object range selection and layer-major assembly at the storage server
before the RDMA transfer.




\subsection{Tracing and Rate Control}

To understand request overhead, the prototype propagates request identifiers
from the NIXL client through RGW into DAOS traces, allowing the evaluation to
split latency into client, gateway, storage, and RDMA-bulk components. For the
multi-tenant scheduling experiments, we add a layerwise pacing control that
limits the effective layer delivery rate. This lets us emulate shared bandwidth
caps while keeping the same layerwise access pattern used by ObjectCache.

\subsection{Testbed}

The evaluation uses a 100~Gbps RoCEv2 (RDMA over Converged Ethernet, v2) cluster with a GPU serving node for inference, an S3 gateway,
and a DAOS storage node. The GPU is an A100 80~GB GPU (CUDA 12.8, PyTorch 2.10.0, vLLM v0.19.0, the modified LMCache v0.4.2, the modified NIXL v1.0.0); the gateway runs the modified Ceph RGW v18.2.7; the storage node runs DAOS v2.7.102 with its default UCX \cite{ucx} transport, equipped with 4 KIOXIA enterprise NVMe SSDs.
In this stack, vLLM is the inference runtime, LMCache manages KV cache reuse on the GPU serving node, NIXL provides the unified storage/memory transfer interface used by both the serving node and the gateway, Ceph RGW terminates S3 requests, DAOS is the underlying object-storage backend, and UCX is the RDMA transport beneath DAOS.
DAOS stripes data across all 4 SSDs; however, because layerwise
range reads access non-contiguous offsets within each chunk object,
the per-SSD access pattern is random rather than sequential.
The measurement of the same micro-benchmark for H100 80GB is in Appendix.

%% file: sections/05_evaluation.tex
\section{Evaluation}
\label{sec:eval}

Our evaluation follows the same mechanisms introduced in the design. First,
the storage and network substrate must transfer data fast enough. Second, the
S3 control path must be measured separately from the RDMA data path. Third,
ObjectCache's server-side aggregation must achieve high throughput for
fine-grained KV chunks. Fourth, layerwise delivery must overlap with GPU
compute in configurations where measured compute and transfer rates make
overlap feasible. Finally, under a shared bandwidth cap, bandwidth-aware
scheduling should minimize TTFT.




\subsection{Raw Storage Baseline}
\label{sec:eval:fio}

\paragraph{Setup.} We measure DAOS throughput as seen by the NIXL
object client without the Ceph RGW gateway, isolating the storage
backend from S3 protocol overhead (Figure \ref{fig:nixl_dfs_baseline}). The benchmark ranges block
sizes from 64~KB to 4~MB and client concurrency
$C \in \{8,32\}$, where $C{=}8$ uses a single thread with 8
in-flight requests and $C{=}32$ uses 4 threads each with 8
in-flight requests.  Writes populate deterministic per-thread key
spaces; a large scrub workload flushes the DAOS cache before
measuring the cold-read throughput.


Local DAOS reads exceed the 100~Gbps NIC capacity at 256 KB block
size with $C{=}32$, and saturate the SSD throughput at 4 MB block size with $C{=}32$.  DAOS over RDMA approaches the 100~Gbps
hardware limit on cold reads at 1 MB block size with $C{=}8$, while TCP
lags consistently (Figure~\ref{fig:nixl_dfs_baseline}).  We use
the saturated RDMA configurations as the reference point for later S3 and
ObjectCache experiments: any remaining gap is attributable to the S3
interface, gateway processing, GPU landing, or aggregation policy
rather than the DAOS backend alone.

\subsection{S3 Transport Baseline}
\label{sec:eval:transport}

\paragraph{Setup.} We next measure S3 PUT and GET throughput across
block sizes from 64~KB to 4~MB. The benchmark compares S3TCP, S3RDMA Buffer that uses a gateway staging buffer, and S3RDMA Direct
where the S3 request carries control information data going through the
DAOS RDMA data path; Section~\ref{sec:impl} gives the corresponding path
names. The two concurrency points, $C{=}8$ and $C{=}32$, match the
canonical raw-storage configurations in Figure~\ref{fig:nixl_dfs_baseline}.
Figure~\ref{fig:s3_throughput} shows that S3RDMA Direct approaches the NIC capacity at 4 MB block size with $C{=}32$. S3TCP
and S3RDMA Buffer suffer from different bottlenecks. S3TCP is limited by the gateway's
streaming HTTP path, while S3RDMA Buffer pays a server-side staging cost. 
Figure~\ref{fig:s3_throughput} also shows why transport alone is not sufficient:
the single-object overhead remains visible for small KV chunks.

\subsection{Layerwise Overlap Model with Hit Rate}
\label{sec:eval:overlap}


What matters for inference is not whether the full prefix arrives
quickly, but whether each layer of cached KV arrives before the GPU
needs it. For a context length $P$ (in tokens) with prefix hit rate $r$ (the
fraction of the $P$ context tokens that are covered by the cached prefix), the
matched KV bytes per layer are
$D^{(\ell)} = 2\,n_{\mathrm{kv}}\,d\,p\,(Pr)$. Given the measured per-layer
compute time $t^{(\ell)}$, perfect overlap requires transfer throughput
$B_{\mathrm{req}} = D^{(\ell)} / t^{(\ell)}$. These expressions connect the
byte layout from Equation~\ref{eq:kv-bytes} to the TTFT model in
Equation~\ref{eq:layerwise-ttft}. If ObjectCache's effective
per-layer throughput exceeds $B_{\mathrm{req}}$, only the
first-layer transfer latency is visible in the TTFT; otherwise, each
layer whose transfer exceeds its compute window adds directly
to the total TTFT.

\begin{figure}[t]
\centering
\includegraphics[width=0.75\columnwidth]{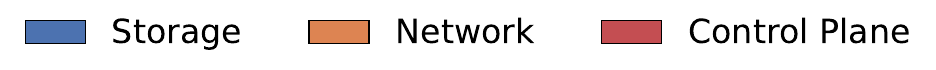}\\[0.2em]
\begin{subfigure}[b]{0.48\columnwidth}
  \centering
  \includegraphics[width=\textwidth]{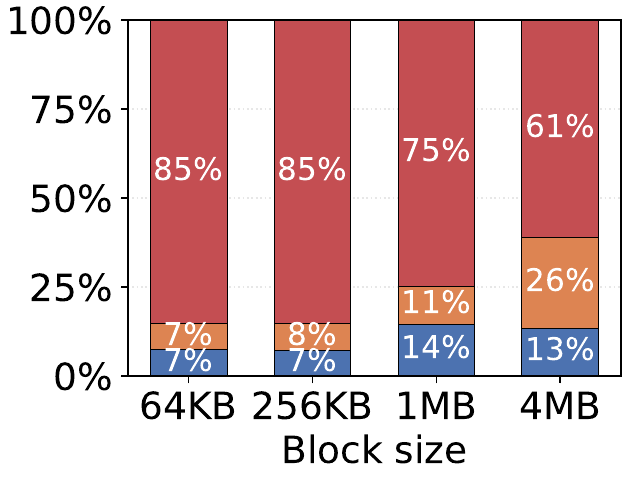}
  \caption{GET.}
  \label{fig:http_overhead:read}
\end{subfigure}%
\hfill
\begin{subfigure}[b]{0.48\columnwidth}
  \centering
  \includegraphics[width=\textwidth]{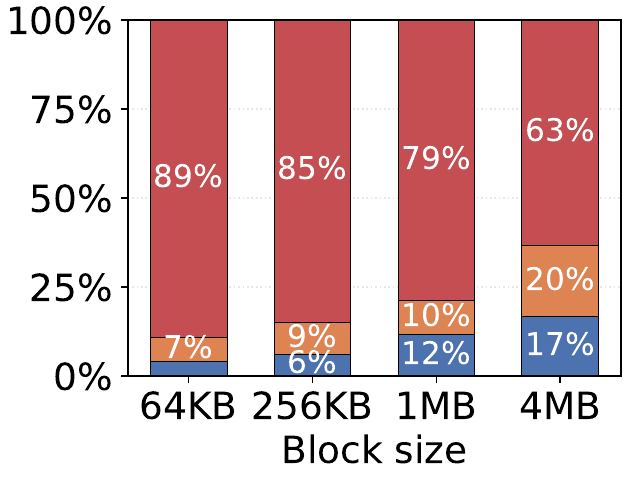}
  \caption{PUT.}
  \label{fig:http_overhead:write}
\end{subfigure}
\caption{Per-request latency breakdown of S3RDMA Direct. Storage is backend
object I/O, Network is RDMA data-plane transfer, and Control Plane is S3
frontend request and metadata processing. For small objects, fixed control-plane
work dominates the remaining latency after RDMA removes TCP data movement.
}
\label{fig:http_overhead}
\end{figure}

\begin{figure}[t]
\centering
\includegraphics[width=0.95\columnwidth]{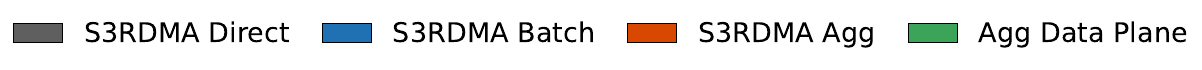}\\[0.2em]
\includegraphics[width=\columnwidth]{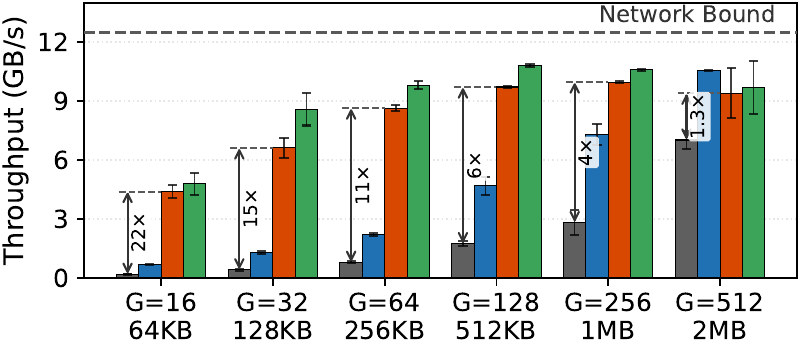}
\caption{Server-side aggregation amortizes per-object overhead and achieves high speedups at small chunk granularities.
}
\label{fig:transport_sweep}
\end{figure}


\subsection{Per-Request Overhead and Overlap Feasibility}
\label{sec:eval:interface}

Figures~\ref{fig:http_overhead} and~\ref{fig:transport_sweep} first isolate the
two interface costs that ObjectCache must address: fixed per-request control cost
and low throughput for fine-grained objects.
After RDMA reduces transfer overhead, HTTP and RGW metadata work
dominate the remaining per-request cost at small objects (Figure~\ref{fig:http_overhead}).
Figure~\ref{fig:transport_sweep} shows that batching and server-side
aggregation turn many small object reads into larger layerwise transfers and
recover high throughput for KV cache loads. Together, these measurements show
why ObjectCache needs both pieces: batching amortizes fixed request overhead, while
aggregation makes fine-grained KV chunks efficient to transfer.

\begin{figure}[t]
\centering
\newcommand{\mainheatmaplegend}[1]{\includegraphics[width=\textwidth]{#1}\\[0.15em]}
\newcommand{\mainheatmap}[1]{\includegraphics[width=\textwidth]{#1}}
\begin{subfigure}[t]{0.48\columnwidth}
  \centering
  \mainheatmaplegend{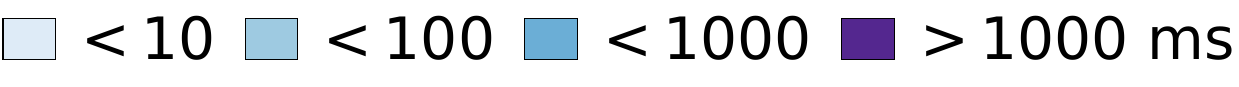}
  \mainheatmap{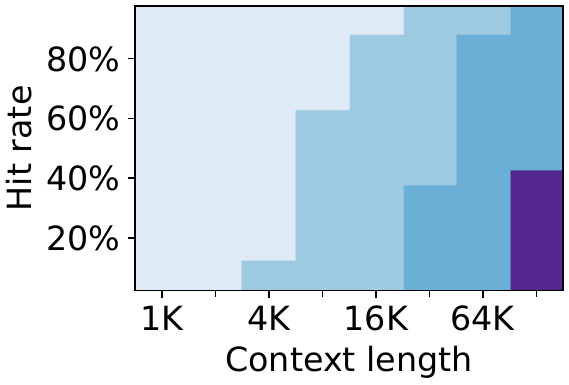}
  \caption{Llama~3.1~8B per-layer compute time.}
  \label{fig:recompute:time}
\end{subfigure}%
\hfill
\begin{subfigure}[t]{0.48\columnwidth}
  \centering
  \mainheatmaplegend{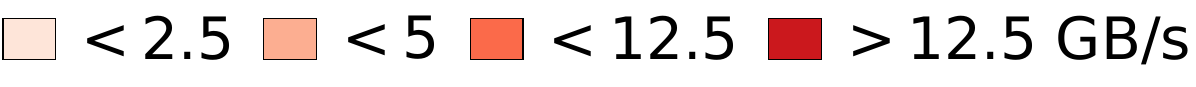}
  \mainheatmap{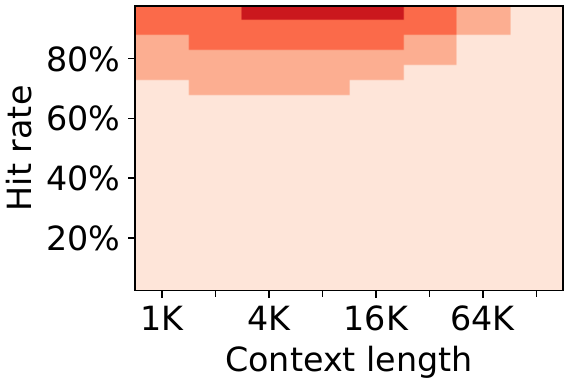}
  \caption{Llama~3.1~8B required transfer throughput to overlap.}
  \label{fig:recompute:bw:a100}
\end{subfigure}
\vspace{0.4em}

\begin{subfigure}[t]{0.48\columnwidth}
  \centering
  \mainheatmaplegend{figures/multimodel_required_bw_heatmap_legend.pdf}
  \mainheatmap{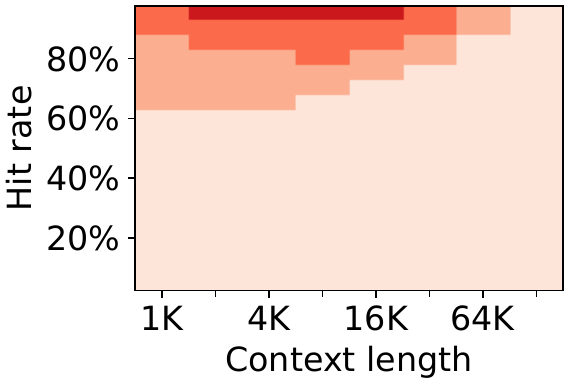}
  \caption{Granite~3.3~8B required transfer throughput to overlap.}
  \label{fig:recompute:bw:a100:granite}
\end{subfigure}%
\hfill
\begin{subfigure}[t]{0.48\columnwidth}
  \centering
  \mainheatmaplegend{figures/multimodel_required_bw_heatmap_legend.pdf}
  \mainheatmap{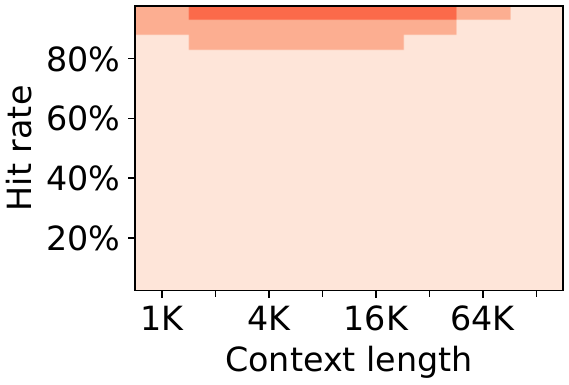}
  \caption{DeepSeek-R1-Distill-Qwen-7B required transfer throughput to overlap.}
  \label{fig:recompute:bw:a100:deepseek}
\end{subfigure}
\caption{A100 layerwise overlap requirements. Most evaluated configurations
require less than 2.5GB/s of layerwise transfer bandwidth, so ObjectCache can hide
cached KV transfer under prefill compute in the dominant regimes.}
\label{fig:recompute_heatmap}
\end{figure}

Figure~\ref{fig:recompute_heatmap} connects aggregation throughput to serving.
The first heatmap reports measured per-layer compute time for Llama~3.1~8B,
while the remaining heatmaps report the transfer throughput required 
for Llama, Granite \cite{granite_3.3_8B}, and
DeepSeek \cite{deepseek_r1_qwen_7B} models. Configurations requiring less bandwidth than the
ObjectCache layer throughput are compute-bound; configurations above that boundary suffer from added latency.
The counter-intuitive takeaway is that longer contexts can relax the transfer
budget: although they carry more cached KV bytes, they also create a larger
per-layer compute window in which ObjectCache can hide layerwise transfer.
ObjectCache also reduces many cross-object range reads per layer to a single RDMA
payload per layer (Appendix Table~\ref{tab:lw_opcount}).




\begin{figure*}[t]
\centering
\includegraphics[width=0.85\textwidth]{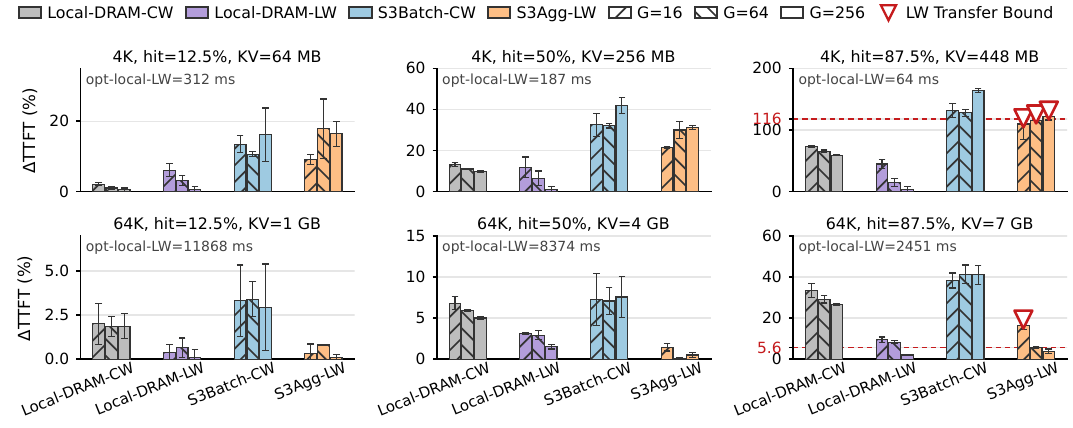}
\caption{TTFT overhead for Llama~3.1~8B relative to the measured optimal local
layerwise baseline for each workload configuration. CW denotes chunkwise
delivery and LW denotes layerwise delivery; S3Agg-LW stays close to the local
baseline except in the transfer-bound corner where layer delivery cannot be
hidden by compute.}
\vspace{+1em}
\label{fig:ttft_delta_over_best}
\end{figure*}

\begin{figure}[t]
\centering
\includegraphics[width=\columnwidth]{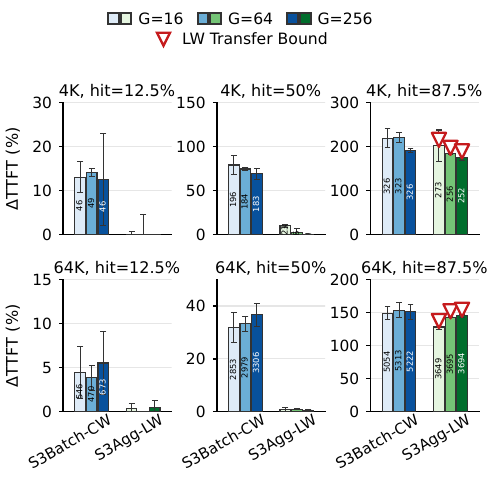}
\caption{Sensitivity of S3-backed KV loading to bandwidth changes for
Llama~3.1~8B. Each bar reports the relative TTFT increase when the same path and
granularity are capped at 10\,Gbps, using its 100\,Gbps result as the baseline.}
\label{fig:ttft_bandwidth_resistance}
\end{figure}



\subsection{End-to-End TTFT}

We evaluate end-to-end TTFT for Llama~3.1~8B across two context
lengths, 4K and 64K, three prefix hit rates, 12.5\%, 50\%, and
87.5\%, and three chunk granularities, $G \in \{16,64,256\}$.
For each workload configuration, the baseline is \textbf{opt-local-LW}: a
pre-aggregated KV cache in layer-major
order in pinned host-memory, so inference requires only host-to-device transfers and incurs
no runtime aggregation cost.
For all other TTFT measurements, we store KV cache data chunkwise with the
specified $G$ in local DRAM or on the remote DAOS server.
Figure~\ref{fig:ttft_delta_over_best} reports the TTFT overhead of
each configuration relative to this baseline.

The results show that layerwise delivery is important regardless of
where the KV cache is stored.
Local-DRAM-LW consistently outperforms Local-DRAM-CW, indicating that
delivering KV blocks in layer order enables better overlap between
transfer and GPU computation.
In contrast, S3Batch-CW incurs the highest overhead in most
configurations because it combines chunkwise delivery with S3
control-plane and network transfer costs.
S3Agg-LW substantially reduces this overhead and performs close to
Local-DRAM-LW in most cases.
In several configurations, S3Agg-LW even achieves lower TTFT than
Local-DRAM-LW.
We interpret this as an observed resource-isolation effect: server-side
aggregation uses dedicated CPU cores and parallel SSD reads, whereas
Local-DRAM-LW consumes client-side CPU and memory bandwidth that may also be
needed by the inference engine.

The benefit of S3Agg-LW is most evident for long-context workloads.
At 64K context length, the per-layer compute window is sufficiently
large to hide most transfer latency, and S3Agg-LW remains within
0.1--5.6\% of opt-local-LW for $G=64$ across all 64K configurations.
Even at the highest hit rate, where the KV payload reaches 7\,GB, the
predicted overlap throughput is 3.1\,GB/s
(Appendix Table~\ref{tab:canonical_cells_compute}), which is below
ObjectCache's sustained aggregation bandwidth.
However, $G=16$ shows noticeably higher overhead than $G=64$ and
$G=256$, suggesting that small chunk granularity prevents the pipeline
from fully exploiting the available server-side aggregation throughput
of approximately 5\,GB/s.

The short-context 4K workloads are more challenging because the compute
window is much smaller.
In these cases, the main $G=64$ S3Agg-LW configuration adds
56--75\,ms over opt-local-LW, and its TTFT can become comparable to
S3Batch-CW.
The limiting case is the 4K, 87.5\% hit-rate workload, where the
baseline compute time is only 64\,ms and the required overlap
throughput rises to 7.4\,GB/s
(Appendix Table~\ref{tab:canonical_cells_compute}).
Although the KV payload is only 448\,MB, fixed costs such as RDMA
session setup, first-layer transfer, and control-plane exchange consume
a significant fraction of the available compute window.
As a result, all three granularities converge to similar TTFT overhead
in this configuration, indicating that the bottleneck is the short compute
window rather than steady-state transfer throughput. We therefore classify all
three granularities in this configuration as transfer-bound: layer delivery, not
steady-state bandwidth efficiency, determines the tail.
Chunk granularity has a modest effect on chunkwise methods.
For Local-DRAM-CW, larger $G$ improves efficiency by reducing
chunk management overhead.
For S3Batch-CW, the effect is weaker because even $G=16$ produces
objects large enough to achieve reasonably high network transfer
efficiency.
These 4K results are consistent with the thresholded mode selection in
Section~\ref{sec:design:mode-selection}: small-payload regimes can be
dominated by fixed and startup costs, so ObjectCache should not assume that
aggregation is always the better delivery mode.

\subsection{Sensitivity to Bandwidth Changes}

We evaluate bandwidth sensitivity by capping the S3-backed paths at
10~Gbps and reporting the TTFT increase relative to the corresponding
100~Gbps run.  Figure~\ref{fig:ttft_bandwidth_resistance} shows that
S3Agg-LW is less sensitive to this bandwidth reduction than
S3Batch-CW except in transfer-bound configurations.

This behavior follows from layerwise overlap.  In S3Batch-CW,
computation cannot start until the chunkwise KV load completes, so a
bandwidth reduction largely appears as additional TTFT.  In S3Agg-LW,
only the non-overlapped portion of each layer transfer contributes to
TTFT.  Therefore, bandwidth changes affect TTFT mainly when the required
overlap throughput exceeds the available bandwidth.
Appendix Table~\ref{tab:canonical_cells_compute} explains where this
happens.  At 64K and 50\% hit rate, the required throughput is only
0.50\,GB/s, well below 10~Gbps, so S3Agg-LW shows almost no sensitivity.
At 4K and 50\% hit rate, the predicted requirement is 1.45\,GB/s,
slightly above 10~Gbps, but the 100~Gbps run is already dominated by
fixed costs and pipeline startup overheads; hence the bandwidth cap does
not visibly dominate end-to-end TTFT.

The large increases appear at 87.5\% hit rate, where the required
throughput exceeds the capped bandwidth: 7.41\,GB/s for 4K and
3.10\,GB/s for 64K.  These configurations become transfer-bound, as indicated by
the markers in Figure~\ref{fig:ttft_bandwidth_resistance}, and therefore
show the largest TTFT increase for S3Agg-LW.  Overall, the results
validate that layerwise loading is intrinsically less sensitive to
bandwidth changes, but only while per-layer transfer can be hidden behind
compute.

\begin{figure}[t]
\centering
\includegraphics[width=\columnwidth]{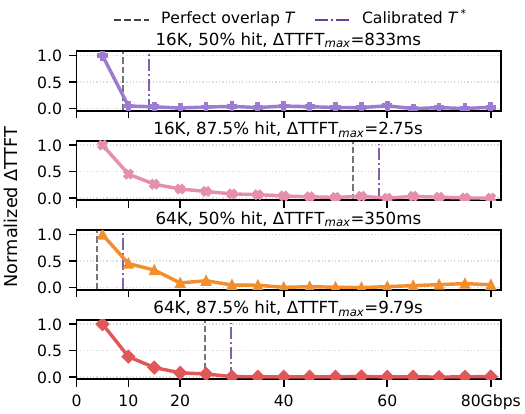}
\caption{Sensitivity of layerwise TTFT to throttled transfer throughput with
S3Agg-LW. Each panel normalizes TTFT increase relative to its best measured
point. Dashed lines show the perfect-overlap bandwidth estimate; dash-dot lines
show the calibrated scheduler target.}
\label{fig:s3agg_lw_rate_sweep}
\end{figure}

\begin{figure}[t]
\centering
\includegraphics[width=0.75\columnwidth]{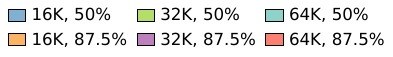}\\[-0.25em]
\begin{subfigure}{0.49\columnwidth}
\centering
\includegraphics[width=\textwidth]{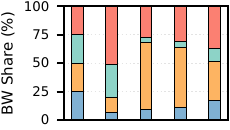}
\caption{Workload-A allocation.}
\end{subfigure}\hfill
\begin{subfigure}{0.49\columnwidth}
\centering
\includegraphics[width=\textwidth]{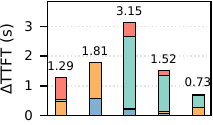}
\caption{Workload-A $\Delta$TTFT.}
\end{subfigure}\\[-0.15em]
\begin{subfigure}{0.49\columnwidth}
\centering
\includegraphics[width=\textwidth]{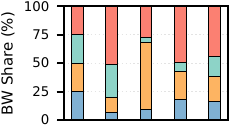}
\caption{Workload-B allocation.}
\end{subfigure}\hfill
\begin{subfigure}{0.49\columnwidth}
\centering
\includegraphics[width=\textwidth]{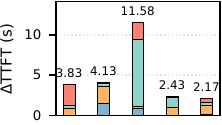}
\caption{Workload-B $\Delta$TTFT.}
\end{subfigure}\\[-0.15em]
\begin{subfigure}{0.49\columnwidth}
\centering
\includegraphics[width=\textwidth]{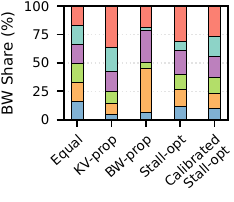}
\caption{Workload-C allocation.}
\end{subfigure}\hfill
\begin{subfigure}{0.49\columnwidth}
\centering
\includegraphics[width=\textwidth]{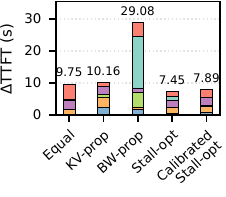}
\caption{Workload-C $\Delta$TTFT.}
\end{subfigure}
\caption{Bandwidth scheduling under shared transfer caps. For each workload,
the left panel shows the bandwidth allocated by each policy and the right panel
shows the resulting added TTFT. Workload-A uses an 80~Gbps cap; Workload-B and
Workload-C use 50~Gbps caps.}
\label{fig:lwcpu_scheduler_methods}
\end{figure}

\subsection{Bandwidth Allocation for Multi Tenants}
\label{sec:bandwidth_allocation_for_multi_tenants}
We next evaluate how the bandwidth allocator behaves when multiple
S3Agg-LW retrievals share a fixed transfer budget. Section~\ref{sec:design:scheduling}
defines Stall-opt and Calibrated Stall-opt; here we compare them against three
heuristic bandwidth-allocation baselines that do not model the layerwise
overlap condition. \emph{Equal}
assigns the same bandwidth to every active request, which is attractive for
fairness but ignores both KV cache size and compute slack. \emph{KV-prop}
allocates bandwidth in proportion to the retrieved KV cache size, favoring
larger or higher-hit-rate requests. \emph{BW-prop} allocates bandwidth in
proportion to the zero-stall bandwidth estimate from
$B_{\mathrm{req}}$, so requests that require higher bandwidth to hide layerwise transfer receive a larger share. For Calibrated Stall-opt,
we use a 5~Gbps margin, chosen from the S3Agg-LW rate sweep in
Figure~\ref{fig:s3agg_lw_rate_sweep}: the dash-dot calibrated targets shift the
analytic perfect-overlap estimates to the measured knees of the TTFT curves.

Figure~\ref{fig:lwcpu_scheduler_methods} compares the scheduling policies on
three mixed workloads, and Appendix
Table~\ref{tab:per_method_allocations_16k64k} reports the exact per-request
bandwidth allocations.
We construct the workloads from the per-request bandwidth requirements in
Appendix Table~\ref{tab:canonical_cells_compute}.
Workloads~A and~B contain the same four requests: 16K and 64K contexts at
50\% and 87.5\% hit rates.
Together, these requests require 91~Gbps in aggregate, or 111~Gbps after
adding the 5~Gbps calibration margin to each request.
The two workloads differ only in their shared bandwidth cap.
Workload~A uses an 80~Gbps cap to model moderate contention, where several
requests can still remain near their calibrated overlap points.
Workload~B uses a 50~Gbps cap to stress the allocator under a tighter budget.
Workload~C extends this setting by adding 32K-context requests, resulting in
six tenants with small, medium, and large KV cache footprints under the same
50~Gbps cap.
Its aggregate requirement is 137~Gbps, or 167~Gbps after calibration.
This workload tests whether the scheduler remains effective when the available
bandwidth must be distributed across a denser set of transfer sensitivities.
The measured per-configuration TTFTs are listed in Appendix
Tables~\ref{tab:s3agg_lw_cap_exact_cells}
and~\ref{tab:s3agg_lw_cap_exact_cells_workload_c}.

Across the three workloads, Calibrated Stall-opt outperforms the heuristic
baselines by allocating bandwidth according to measured overlap requirements
rather than per-request parity, KV size, or analytic demand alone.
Measured as additional TTFT over the effectively unthrottled baseline, it
reduces Equal's penalty by 1.2--1.8$\times$
(Figure~\ref{fig:lwcpu_scheduler_methods}); Appendix
Table~\ref{tab:s3agg_lw_cap_total_ttft} gives the full breakdown.
This improvement comes from assigning bandwidth near each request's useful
overlap region.
Equal ignores heterogeneous transfer sensitivity, KV-prop can starve small but
latency-sensitive requests, and BW-prop can over-allocate beyond the point where
extra bandwidth reduces TTFT.
Calibrated Stall-opt avoids these extremes and reduces added TTFT under both
moderate and tight caps.
The only exception is Workload~C, where Stall-opt is slightly better because the
calibration margin can mildly over-provision some requests under a dense,
heterogeneous 50~Gbps cap.
Even there, Calibrated Stall-opt remains better than Equal, KV-prop, and
BW-prop, preserving most of Stall-opt's benefit while improving robustness to
measurement-level deviations from the analytic model.

%% file: sections/06_discussion.tex
\section{Discussion}
\label{sec:discussion}

\subsection{Position in the KV cache Hierarchy}

ObjectCache is not a replacement for hot KV-cache tier, where GPU memory or DRAM-backed KV cache pools
serve those hottest prefixes.
ObjectCache instead provides a persistent, shared, S3-compatible capacity tier for
large prefix pools, where placement flexibility and low storage cost matter.
The key question is whether
it can be made fast enough for serving-path prefix reuse.
This role complements disaggregated inference systems that separate prefill,
decode, and KV cache storage
\cite{zhong2024distserve,patel2024splitwise,qin2024mooncake,hu2024memserve}.
By storing long-lived prefixes in object storage, ObjectCache lets requests run on
any available GPU serving node rather than only where the prefix is cached.

\subsection{When Layerwise Object Storage Helps}

Layerwise ObjectCache is most useful when recomputation is costly, partial prefill computation provides enough per-layer overlap window, and the storage tier can
deliver near the required overlap bandwidth.
When the compute window is too small, bandwidth is below the overlap
requirement, or the prefix hit is too small to amortize S3 overheads, the system
should instead fall back to chunkwise transfer, a DRAM hot cache, or
recomputation.

\subsection{Limitations and Future Work}
In this paper we optimize the serving-path data movement mechanism for S3-compatible object storage.
Our prototype evaluates the critical data path and scheduling policy on a small
cluster, which is sufficient to isolate the effects of aggregation, RDMA
transfer, layerwise delivery, and bandwidth scheduling, but does not capture all
production effects.
Future work should validate ObjectCache with concurrent clients, long-running
prefix churn, failures, and integration with batching \cite{zheng2025batchllm}, chunked
prefill \cite{agrawal2025efficient}, routing \cite{ding2024hybrid}, and eviction policies \cite{wang2025kvcache, feng2026ada}.

ObjectCache also assumes that cached KV can be addressed by prefix chunk and layer
range.
Irregular layouts, packed objects, or compressed KV representations \cite{chang2025palu, liu2024cachegen} are
compatible only if the storage server has sufficient metadata to translate
layerwise requests.
Finally, production deployments need scalable connection management,
per-tenant admission control, and stronger isolation around RDMA credentials
\cite{lou2024harmonic,rothenberger2021redmark,tsai2019pythia,simpson2020securing,zhao2025white}.

%% file: sections/07_related_work.tex
\section{Related Work}
\label{sec:related}

\textbf{KV cache management and prefix reuse.}
Paged KV allocation, radix-tree prefix matching, and KV offloading systems
provide the software substrate that ObjectCache builds on
\cite{kwon2023efficient,ye2024chunkattention,sheng2023flexgen,lee2024infinigen,chen2025impress,yao2025cacheblend,yu202510x,zou2026contiguouskv,prabhu2025vattention}.
These systems typically store KV in fixed-size chunks.
ObjectCache preserves this chunkwise, hash-addressed layout, but changes how remote
storage delivers matched chunks to the serving node.
CacheFlow~\cite{nian2026cacheflow} accelerates KV cache restoration by
scheduling recomputation and I/O across token, layer, and GPU dimensions
inside the serving system.
ObjectCache addresses a different bottleneck: how an S3-compatible object tier
should fetch, aggregate, and deliver many matched KV chunks in layer order to
the inference engine.

\textbf{Disaggregated LLM serving.}
Prior work on prefill/decode separation and remote KV pools shows that LLM
serving can benefit from decoupling compute roles and moving KV cache data over the
network
\cite{zhong2024distserve,patel2024splitwise,qin2024mooncake,hu2024memserve,hu2025shuffleinfer,hu2025deepserve,qin2026prefill,chen2026towards,cheng2026kunserve,fu2024serverlessllm,mei2025helix,su2025efficient,kamath2025pod}.
ObjectCache follows this architectural direction, but targets a persistent
S3-compatible object storage which targets for a cloud native solution.


\textbf{S3-over-RDMA and object storage.}
Recent S3-over-RDMA systems split the HTTP control path from an RDMA data path, improving large-object transfer over TCP-based S3
\cite{cuobject,cuobject_server,minio_s3_rdma,nvidia_s3_over_rdma,vast_s3_over_rdma,cloudian_s3_over_rdma,dell_s3_over_rdma,sun2025fast,hao2026fast}.
Low-latency object stores such as S3 Express reduce access latency but still expose single-object request semantics
\cite{zhou2026milliscale}.
ObjectCache builds on this trend by adding KV-aware multi-object aggregation and layerwise delivery to S3-compatible storage.

\textbf{KV compression and small-state attention.}
KV quantization, compression, MLA-style architectures, recent-window methods,
and state-retention techniques reduce the bytes retained per token
\cite{shazeer2019fast,ainslie2023gqa,liu2024deepseek,zhang2023h2o,liu2023scissorhands,xiao2024efficient,li2024snapkv,ge2024model,liu2024kivi,hooper2024kvquant,sunshadowkv,feng2026ada,zandieh2026turboquant,du2026bitdecoding}.
These techniques are complementary to ObjectCache.
They can reduce transfer volume, but they also shrink each useful object
payload, making S3 overhead more dominant.

%% file: sections/08_conclusion.tex
\section{Conclusion}
\label{sec:conclusion}

S3-compatible object storage is a natural storage tier for large prefix
KV cache pools, but standard S3 exposes the mismatched serving-path semantics.
ObjectCache adds the missing serving-path semantics via protocol-scheduling co-design.
Our prototype shows that these pieces must work together. RDMA improves raw S3
transport, aggregation amortizes network overhead, layerwise delivery
overlaps transfer with GPU compute, and bandwidth-aware scheduling reduces
added TTFT under shared bandwidth.
On a 100~Gbps RoCE prototype with Llama 3.1 8B, ObjectCache reaches within 5.6\%
of the optimal local DRAM solution at 64K context length.
Our results suggest that object storage can serve as a runtime KV cache
backend rather than only a cold archive, while retaining the persistence,
capacity, placement flexibility, and low storage cost that make object storage
attractive.




%% file: sections/09_appendix.tex
\appendix
\renewcommand{\thefigure}{A\arabic{figure}}
\renewcommand{\thetable}{A\arabic{table}}
\setcounter{figure}{0}
\setcounter{table}{0}

\section{Additional Motivation Data}
\label{app:motivation-data}

Tables~\ref{tab:boundary_granularity_llama} and
\ref{tab:boundary_granularity_deepseek} report the full steady-state
measurements behind the cache-boundary granularity study. For each semantic
hit boundary, the prompt reuses \(M-G\) tokens, where \(M=C\cdot r\) is
the base hit boundary and \(G\in\{16,512\}\). Thus \(G=512\) recomputes
496 more tokens than \(G=16\) in every row. Entries are TTFT in milliseconds
over steady-state trials 1--2, reported as mean \(\pm\) one standard
deviation. \(\Delta = T_{G=512}-T_{G=16}\) uses the same steady-state
trials and excludes trial~0 to avoid shape-warmup effects.

\begin{table}[t]
\centering
\caption{Steady-state boundary-granularity recompute cost for Llama~3.1~8B.}
\label{tab:boundary_granularity_llama}
\small
\setlength{\tabcolsep}{2.6pt}
\renewcommand{\arraystretch}{1.04}
\begin{tabular}{llrrrr}
\toprule
\rowcolor{headerblue}
GPU & Ctx. & Hit (\%) & \(G=16\) & \(G=512\) & \(\Delta\) \\
\midrule
A100 & 4K & 12.5 & 322.6 \(\pm\) 8.6 & 344.2 \(\pm\) 5.9 & 21.6 \\
 &  & 25.0 & 270.5 \(\pm\) 0.4 & 310.3 \(\pm\) 1.8 & 39.7 \\
 &  & 37.5 & 236.3 \(\pm\) 1.4 & 264.5 \(\pm\) 0.6 & 28.2 \\
 &  & 50.0 & 192.6 \(\pm\) 0.4 & 222.7 \(\pm\) 0.2 & 30.2 \\
 &  & 62.5 & 149.3 \(\pm\) 2.0 & 182.1 \(\pm\) 0.0 & 32.9 \\
 &  & 75.0 & 109.0 \(\pm\) 0.1 & 141.8 \(\pm\) 1.5 & 32.7 \\
 &  & 87.5 & 70.2 \(\pm\) 0.9 & 101.9 \(\pm\) 0.3 & 31.7 \\
 & 64K & 12.5 & 11643.8 \(\pm\) 9.3 & 11686.9 \(\pm\) 12.7 & 43.1 \\
 &  & 25.0 & 10768.3 \(\pm\) 13.2 & 10810.8 \(\pm\) 1.1 & 42.5 \\
 &  & 37.5 & 9616.7 \(\pm\) 2.3 & 9671.8 \(\pm\) 6.2 & 55.2 \\
 &  & 50.0 & 8236.6 \(\pm\) 6.8 & 8283.8 \(\pm\) 6.5 & 47.2 \\
 &  & 62.5 & 6577.9 \(\pm\) 9.0 & 6645.4 \(\pm\) 9.5 & 67.6 \\
 &  & 75.0 & 4676.2 \(\pm\) 1.0 & 4763.2 \(\pm\) 0.1 & 87.0 \\
 &  & 87.5 & 2538.5 \(\pm\) 3.1 & 2642.0 \(\pm\) 0.5 & 103.6 \\
\midrule
H100 & 4K & 12.5 & 143.5 \(\pm\) 0.2 & 162.3 \(\pm\) 0.5 & 18.8 \\
 &  & 25.0 & 122.8 \(\pm\) 0.0 & 140.7 \(\pm\) 0.3 & 17.9 \\
 &  & 37.5 & 107.1 \(\pm\) 1.5 & 125.9 \(\pm\) 0.5 & 18.8 \\
 &  & 50.0 & 87.5 \(\pm\) 0.7 & 106.2 \(\pm\) 1.5 & 18.6 \\
 &  & 62.5 & 69.0 \(\pm\) 0.1 & 87.9 \(\pm\) 0.9 & 18.9 \\
 &  & 75.0 & 50.0 \(\pm\) 0.1 & 70.8 \(\pm\) 1.0 & 20.8 \\
 &  & 87.5 & 33.7 \(\pm\) 0.1 & 51.7 \(\pm\) 0.8 & 17.9 \\
 & 64K & 12.5 & 4948.6 \(\pm\) 14.4 & 5009.5 \(\pm\) 2.9 & 60.9 \\
 &  & 25.0 & 4547.8 \(\pm\) 0.4 & 4605.5 \(\pm\) 5.5 & 57.7 \\
 &  & 37.5 & 4015.1 \(\pm\) 1.9 & 4067.1 \(\pm\) 2.9 & 52.0 \\
 &  & 50.0 & 3410.1 \(\pm\) 3.6 & 3482.9 \(\pm\) 1.2 & 72.7 \\
 &  & 62.5 & 2699.0 \(\pm\) 0.6 & 2757.0 \(\pm\) 0.7 & 58.0 \\
 &  & 75.0 & 1911.2 \(\pm\) 3.4 & 1985.0 \(\pm\) 3.1 & 73.7 \\
 &  & 87.5 & 1022.1 \(\pm\) 1.1 & 1090.5 \(\pm\) 1.8 & 68.5 \\
\bottomrule
\end{tabular}
\end{table}

\begin{table}[t]
\centering
\caption{Steady-state boundary-granularity recompute cost for DeepSeek-R1-Distill-Qwen-7B.}
\label{tab:boundary_granularity_deepseek}
\small
\setlength{\tabcolsep}{2.6pt}
\renewcommand{\arraystretch}{1.04}
\begin{tabular}{llrrrr}
\toprule
\rowcolor{headerblue}
GPU & Ctx. & Hit (\%) & \(G=16\) & \(G=512\) & \(\Delta\) \\
\midrule
A100 & 4K & 12.5 & 283.1 \(\pm\) 1.2 & 317.0 \(\pm\) 2.8 & 34.0 \\
 &  & 25.0 & 251.2 \(\pm\) 2.9 & 278.5 \(\pm\) 0.3 & 27.3 \\
 &  & 37.5 & 212.0 \(\pm\) 2.9 & 243.7 \(\pm\) 2.9 & 31.7 \\
 &  & 50.0 & 177.7 \(\pm\) 1.6 & 206.3 \(\pm\) 0.4 & 28.6 \\
 &  & 62.5 & 140.6 \(\pm\) 0.7 & 168.5 \(\pm\) 0.9 & 27.8 \\
 &  & 75.0 & 102.6 \(\pm\) 0.1 & 132.2 \(\pm\) 0.3 & 29.6 \\
 &  & 87.5 & 60.2 \(\pm\) 0.0 & 93.5 \(\pm\) 0.3 & 33.3 \\
 & 64K & 12.5 & 9633.1 \(\pm\) 19.9 & 9708.1 \(\pm\) 17.7 & 75.0 \\
 &  & 25.0 & 8893.6 \(\pm\) 26.1 & 8966.9 \(\pm\) 0.1 & 73.3 \\
 &  & 37.5 & 7932.3 \(\pm\) 16.4 & 7994.1 \(\pm\) 8.8 & 61.8 \\
 &  & 50.0 & 6769.8 \(\pm\) 16.1 & 6836.8 \(\pm\) 8.2 & 66.9 \\
 &  & 62.5 & 5414.7 \(\pm\) 2.2 & 5479.2 \(\pm\) 1.3 & 64.5 \\
 &  & 75.0 & 3847.8 \(\pm\) 3.2 & 3914.7 \(\pm\) 19.3 & 66.9 \\
 &  & 87.5 & 2098.0 \(\pm\) 3.2 & 2179.5 \(\pm\) 9.9 & 81.5 \\
\midrule
H100 & 4K & 12.5 & 133.7 \(\pm\) 0.8 & 149.8 \(\pm\) 0.5 & 16.1 \\
 &  & 25.0 & 117.1 \(\pm\) 0.2 & 134.3 \(\pm\) 1.0 & 17.3 \\
 &  & 37.5 & 98.4 \(\pm\) 0.3 & 115.8 \(\pm\) 0.6 & 17.4 \\
 &  & 50.0 & 83.9 \(\pm\) 2.0 & 99.5 \(\pm\) 0.9 & 15.6 \\
 &  & 62.5 & 68.1 \(\pm\) 1.8 & 81.1 \(\pm\) 0.1 & 12.9 \\
 &  & 75.0 & 45.6 \(\pm\) 0.5 & 64.2 \(\pm\) 0.7 & 18.6 \\
 &  & 87.5 & 31.7 \(\pm\) 0.3 & 46.2 \(\pm\) 0.2 & 14.5 \\
 & 64K & 12.5 & 4259.2 \(\pm\) 8.6 & 4311.1 \(\pm\) 0.8 & 51.9 \\
 &  & 25.0 & 3898.4 \(\pm\) 1.2 & 3934.0 \(\pm\) 3.1 & 35.6 \\
 &  & 37.5 & 3425.6 \(\pm\) 0.4 & 3454.9 \(\pm\) 2.3 & 29.3 \\
 &  & 50.0 & 2895.8 \(\pm\) 0.0 & 2922.4 \(\pm\) 2.2 & 26.6 \\
 &  & 62.5 & 2280.2 \(\pm\) 2.8 & 2329.3 \(\pm\) 4.9 & 49.1 \\
 &  & 75.0 & 1616.2 \(\pm\) 0.5 & 1647.2 \(\pm\) 3.0 & 31.0 \\
 &  & 87.5 & 847.2 \(\pm\) 1.1 & 904.5 \(\pm\) 0.2 & 57.2 \\
\bottomrule
\end{tabular}
\end{table}

\newpage
\section{Measured Cache-Boundary Recompute}
\label{app:boundary-granularity-recompute}

\begin{figure}[t]
    \centering
    \includegraphics[width=0.95\columnwidth]{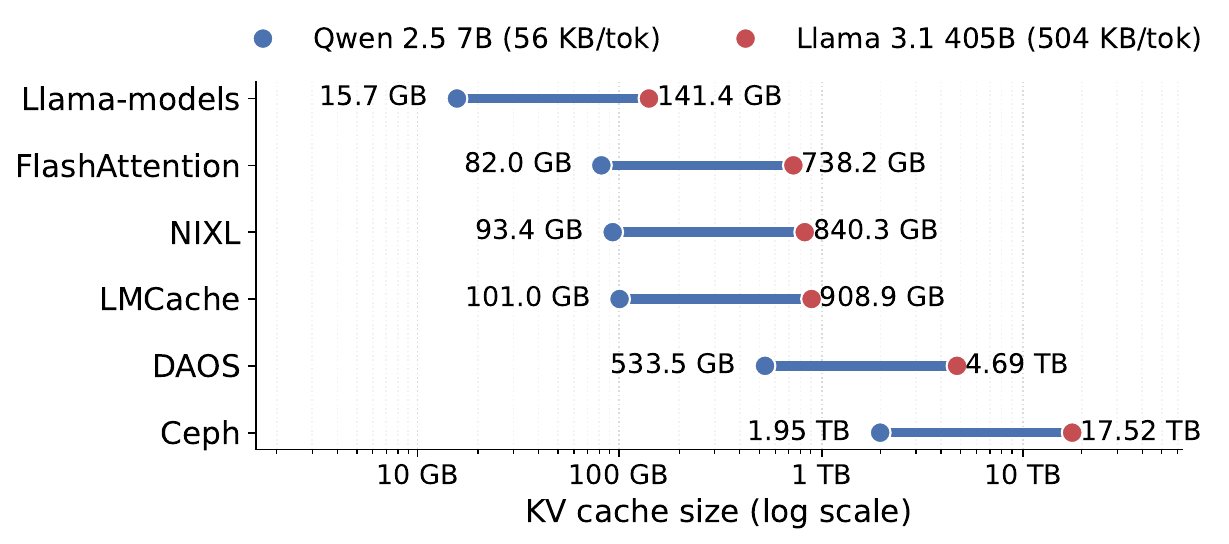}
    \caption{Repository-scale prefixes can translate into large KV cache
    footprints. Ranges are computed from repository token counts and two
    representative KV cache sizes per token.}
    \label{fig:repo_kv_range}
\end{figure}

\begin{figure}[t]
    \centering
    \includegraphics[width=0.95\columnwidth]{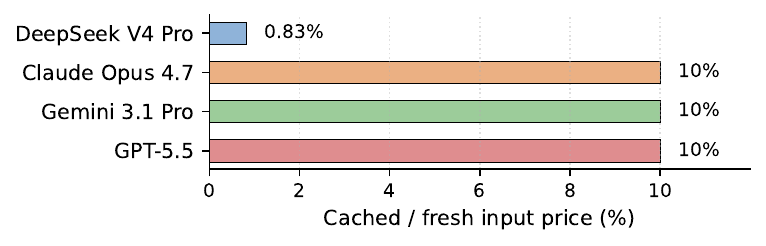}
    \caption{Major APIs price cached input far below fresh input, reflecting
    the economic value of prefix reuse in deployed LLM
    services~\cite{deepseek_price,claude_price,gemini_price,chatgpt_price}.}
    \label{fig:provider_cache_price}
\end{figure}


\section{Pinned CPU-GPU H2D Microbenchmarks}
\label{app:h100-h2d-microbench}

\begin{figure*}[t]
    \centering
    \includegraphics[width=0.75\textwidth]{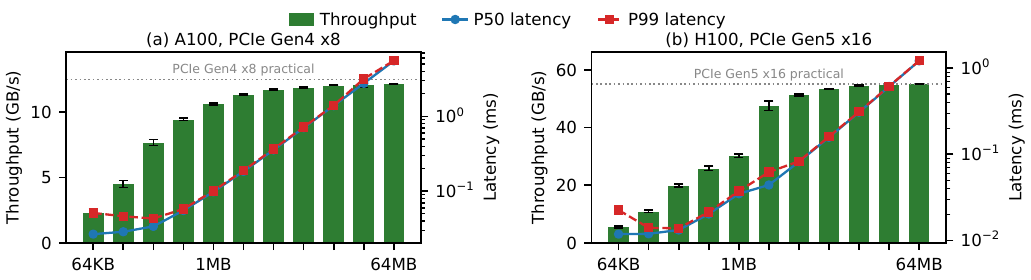}
    \caption{Pinned CPU$\to$GPU H2D throughput and latency vs. block size for
    A100 (PCIe Gen4 x8) and H100 (PCIe Gen5 x16). Bars report mean throughput
    ($\pm$1$\sigma$ over 200 per-size samples); right-axis lines report P50/P99
    latency across 64~KB--64~MB blocks. The source buffer is placed on NUMA~0
    in both setups. A100 saturates near 12~GB/s; H100 saturates near 55~GB/s for
    blocks $\geq$8~MB.}
    \label{fig:h2d_microbench_a100}
\end{figure*}




\section{Per-model, per-GPU recompute and bandwidth heatmaps}
\label{app:multimodel-heatmaps}

\begin{figure*}[t]
    \centering
    \captionsetup[subfigure]{font=scriptsize}
    \includegraphics[width=0.35\textwidth]{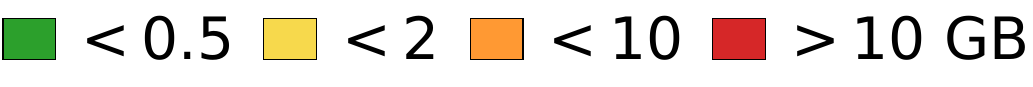}\\[0.4em]
    \begin{subfigure}[t]{0.235\textwidth}
      \centering
      \includegraphics[width=\textwidth]{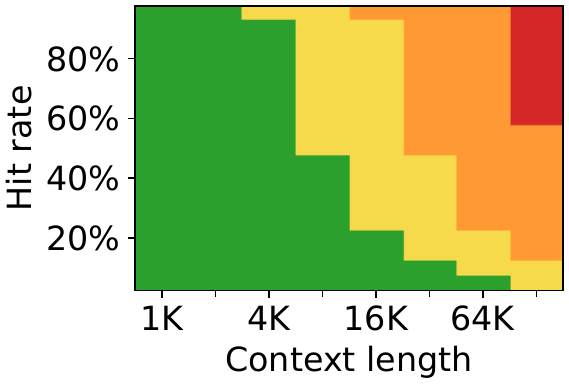}
      \caption{Llama~3.1~8B}
    \end{subfigure}\hfill
    \begin{subfigure}[t]{0.235\textwidth}
      \centering
      \includegraphics[width=\textwidth]{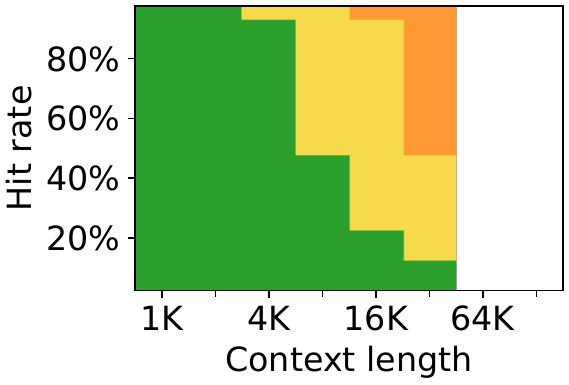}
      \caption{Mistral~7B}
    \end{subfigure}\hfill
    \begin{subfigure}[t]{0.235\textwidth}
      \centering
      \includegraphics[width=\textwidth]{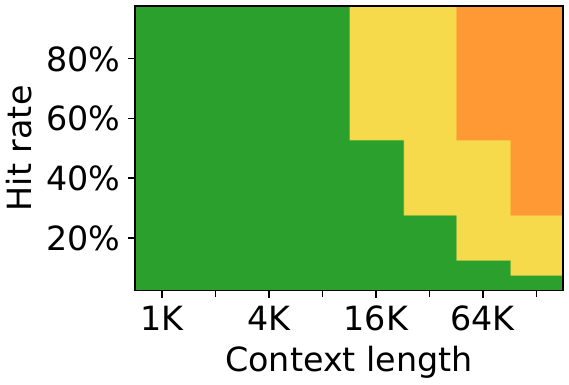}
      \caption{DeepSeek-Distill 7B}
    \end{subfigure}\hfill
    \begin{subfigure}[t]{0.235\textwidth}
      \centering
      \includegraphics[width=\textwidth]{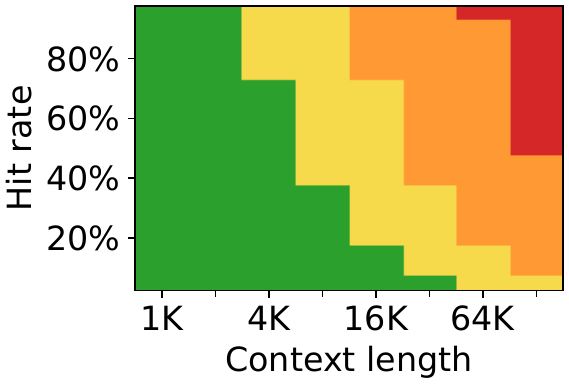}
      \caption{Granite~3.3~8B}
    \end{subfigure}
    \caption{Total cached KV size per (context length, hit rate) configuration
    for each model. KV size is independent of GPU platform, so each value is
    reported once.}
    \label{fig:multimodel_kvcache_size}
\end{figure*}

\begin{figure*}[t]
    \centering
    \captionsetup[subfigure]{font=scriptsize}
    \includegraphics[width=0.45\textwidth]{figures/multimodel_compute_heatmap_legend.pdf}\\[0.4em]
    \begin{subfigure}[t]{0.235\textwidth}
      \centering
      \includegraphics[width=\textwidth]{figures/multimodel_a100_llama_compute_heatmap.pdf}
      \caption{A100, Llama~3.1~8B}
    \end{subfigure}\hfill
    \begin{subfigure}[t]{0.235\textwidth}
      \centering
      \includegraphics[width=\textwidth]{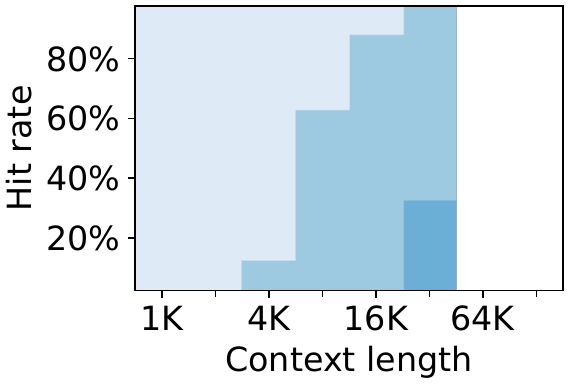}
      \caption{A100, Mistral~7B}
    \end{subfigure}\hfill
    \begin{subfigure}[t]{0.235\textwidth}
      \centering
      \includegraphics[width=\textwidth]{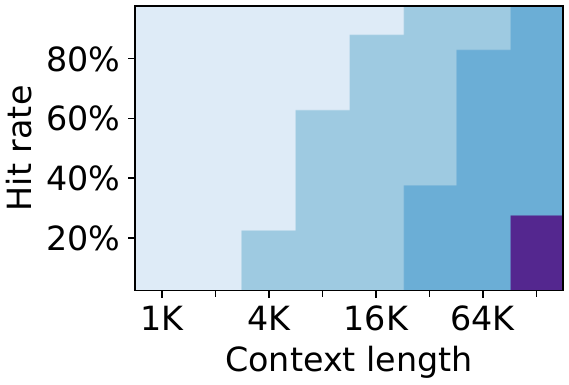}
      \caption{A100, DeepSeek-Distill 7B}
    \end{subfigure}\hfill
    \begin{subfigure}[t]{0.235\textwidth}
      \centering
      \includegraphics[width=\textwidth]{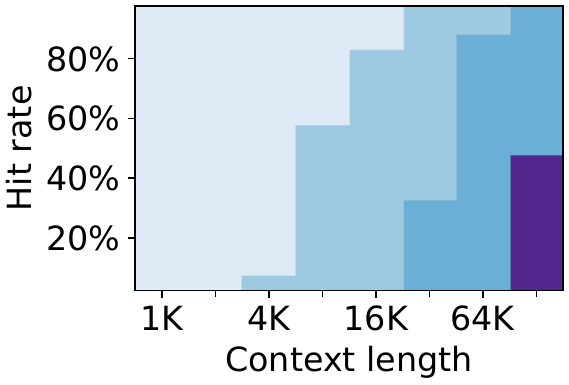}
      \caption{A100, Granite~3.3~8B}
    \end{subfigure}\\[0.4em]
    \begin{subfigure}[t]{0.235\textwidth}
      \centering
      \includegraphics[width=\textwidth]{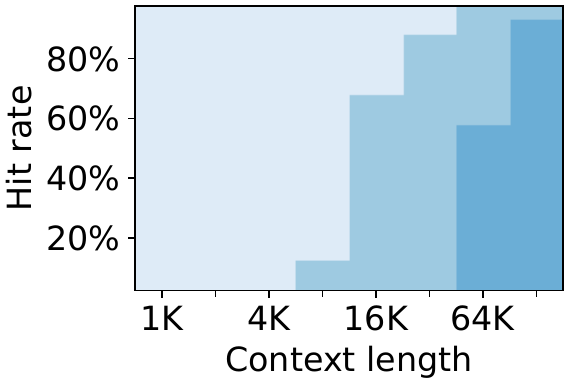}
      \caption{H100, Llama~3.1~8B}
    \end{subfigure}\hfill
    \begin{subfigure}[t]{0.235\textwidth}
      \centering
      \includegraphics[width=\textwidth]{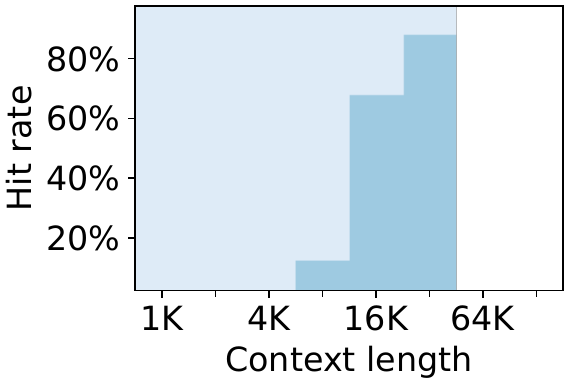}
      \caption{H100, Mistral~7B}
    \end{subfigure}\hfill
    \begin{subfigure}[t]{0.235\textwidth}
      \centering
      \includegraphics[width=\textwidth]{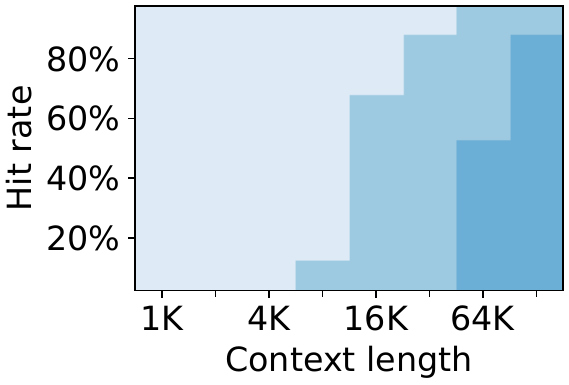}
      \caption{H100, DeepSeek-Distill 7B}
    \end{subfigure}\hfill
    \begin{subfigure}[t]{0.235\textwidth}
      \centering
      \includegraphics[width=\textwidth]{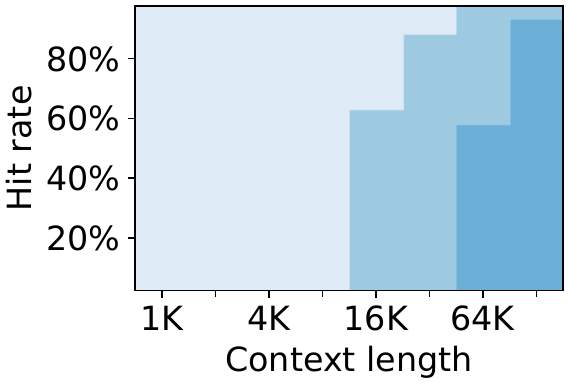}
      \caption{H100, Granite~3.3~8B}
    \end{subfigure}
    \caption{Per-layer compute time across four representative models on A100
    (top row) and H100 (bottom row). Blank configurations indicate unsupported
    or unmeasured context lengths, such as Mistral~7B beyond its 32K window.}
    \label{fig:multimodel_compute_time}
\end{figure*}

\begin{figure*}[t]
    \centering
    \captionsetup[subfigure]{font=scriptsize}
    \includegraphics[width=0.35\textwidth]{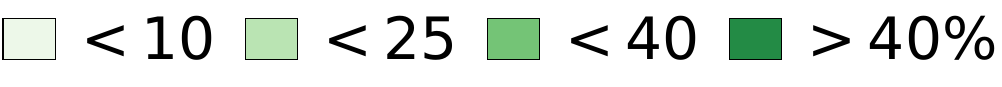}\\[0.4em]
    \begin{subfigure}[t]{0.235\textwidth}
      \centering
      \includegraphics[width=\textwidth]{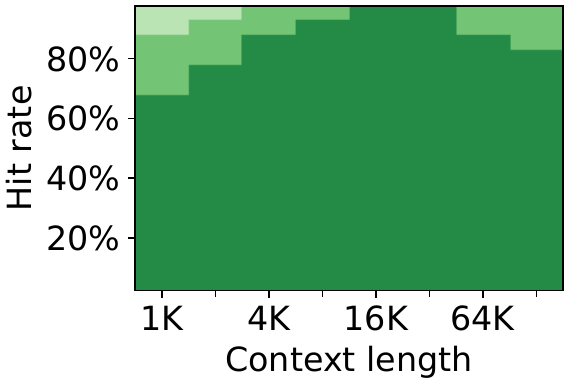}
      \caption{A100, Llama~3.1~8B}
    \end{subfigure}\hfill
    \begin{subfigure}[t]{0.235\textwidth}
      \centering
      \includegraphics[width=\textwidth]{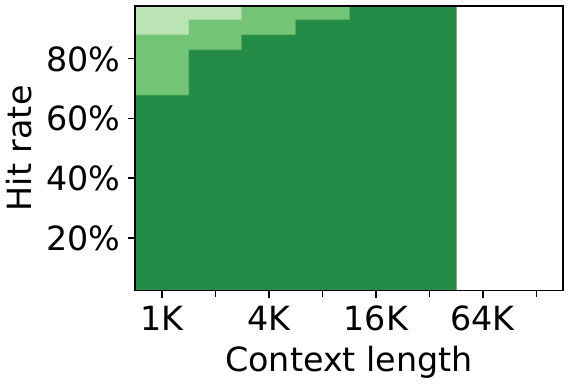}
      \caption{A100, Mistral~7B}
    \end{subfigure}\hfill
    \begin{subfigure}[t]{0.235\textwidth}
      \centering
      \includegraphics[width=\textwidth]{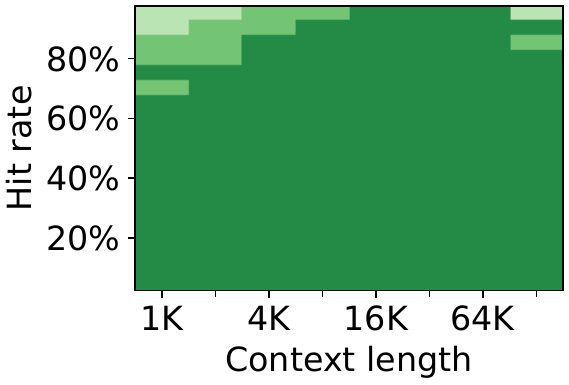}
      \caption{A100, DeepSeek-Distill 7B}
    \end{subfigure}\hfill
    \begin{subfigure}[t]{0.235\textwidth}
      \centering
      \includegraphics[width=\textwidth]{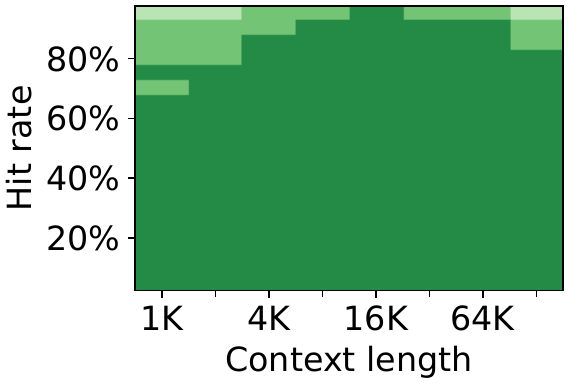}
      \caption{A100, Granite~3.3~8B}
    \end{subfigure}\\[0.4em]
    \begin{subfigure}[t]{0.235\textwidth}
      \centering
      \includegraphics[width=\textwidth]{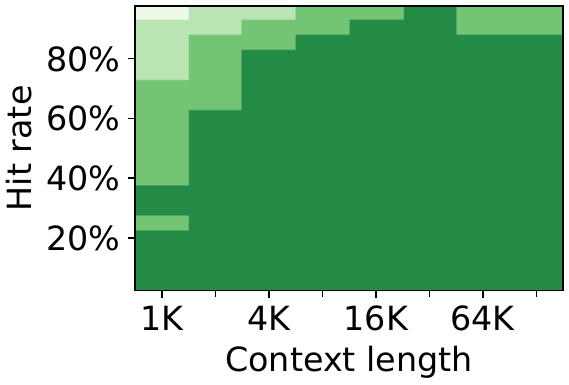}
      \caption{H100, Llama~3.1~8B}
    \end{subfigure}\hfill
    \begin{subfigure}[t]{0.235\textwidth}
      \centering
      \includegraphics[width=\textwidth]{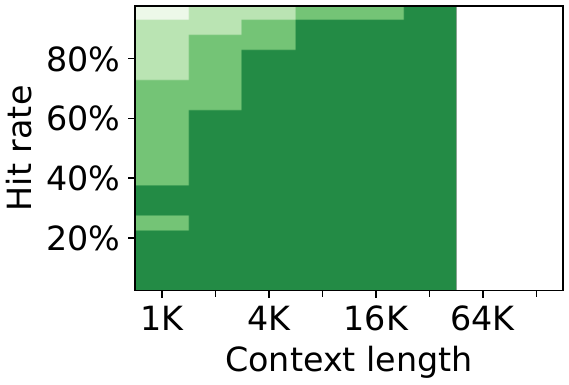}
      \caption{H100, Mistral~7B}
    \end{subfigure}\hfill
    \begin{subfigure}[t]{0.235\textwidth}
      \centering
      \includegraphics[width=\textwidth]{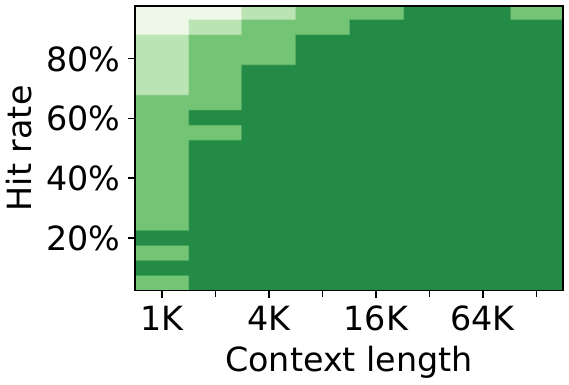}
      \caption{H100, DeepSeek-Distill 7B}
    \end{subfigure}\hfill
    \begin{subfigure}[t]{0.235\textwidth}
      \centering
      \includegraphics[width=\textwidth]{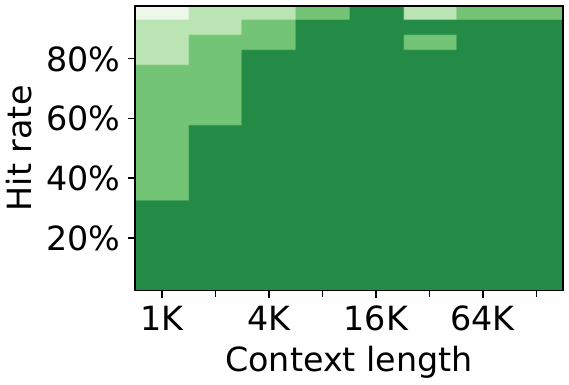}
      \caption{H100, Granite~3.3~8B}
    \end{subfigure}
    \caption{Model FLOP utilization (MFU) across four representative models on
    A100 (top row) and H100 (bottom row). Low-MFU cells indicate configurations
    that do not saturate GPU compute; high-MFU cells identify compute-dense
    configurations.}
    \label{fig:multimodel_mfu}
\end{figure*}

\begin{figure*}[t]
    \centering
    \captionsetup[subfigure]{font=scriptsize}
    \includegraphics[width=0.4\textwidth]{figures/multimodel_required_bw_heatmap_legend.pdf}\\[0.4em]
    \begin{subfigure}[t]{0.235\textwidth}
      \centering
      \includegraphics[width=\textwidth]{figures/multimodel_a100_llama_required_bw_heatmap.pdf}
      \caption{A100, Llama~3.1~8B}
    \end{subfigure}\hfill
    \begin{subfigure}[t]{0.235\textwidth}
      \centering
      \includegraphics[width=\textwidth]{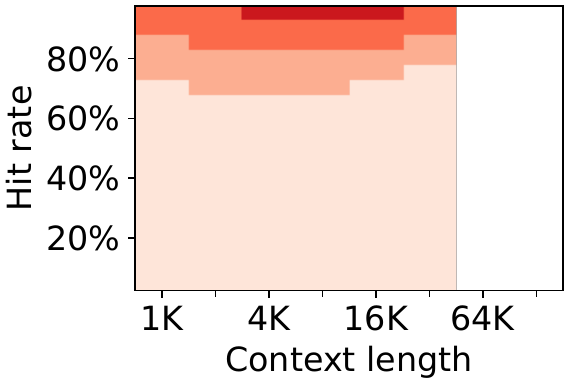}
      \caption{A100, Mistral~7B}
    \end{subfigure}\hfill
    \begin{subfigure}[t]{0.235\textwidth}
      \centering
      \includegraphics[width=\textwidth]{figures/multimodel_a100_deepseek_required_bw_heatmap.pdf}
      \caption{A100, DeepSeek-Distill 7B}
    \end{subfigure}\hfill
    \begin{subfigure}[t]{0.235\textwidth}
      \centering
      \includegraphics[width=\textwidth]{figures/multimodel_a100_granite_required_bw_heatmap.pdf}
      \caption{A100, Granite~3.3~8B}
    \end{subfigure}\\[0.4em]
    \begin{subfigure}[t]{0.235\textwidth}
      \centering
      \includegraphics[width=\textwidth]{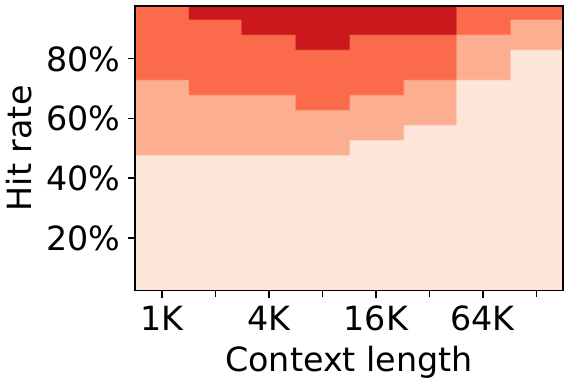}
      \caption{H100, Llama~3.1~8B}
    \end{subfigure}\hfill
    \begin{subfigure}[t]{0.235\textwidth}
      \centering
      \includegraphics[width=\textwidth]{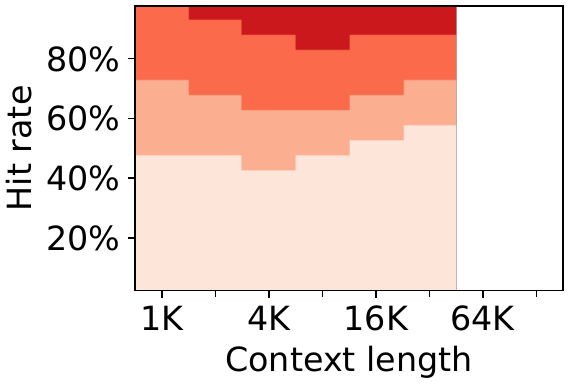}
      \caption{H100, Mistral~7B}
    \end{subfigure}\hfill
    \begin{subfigure}[t]{0.235\textwidth}
      \centering
      \includegraphics[width=\textwidth]{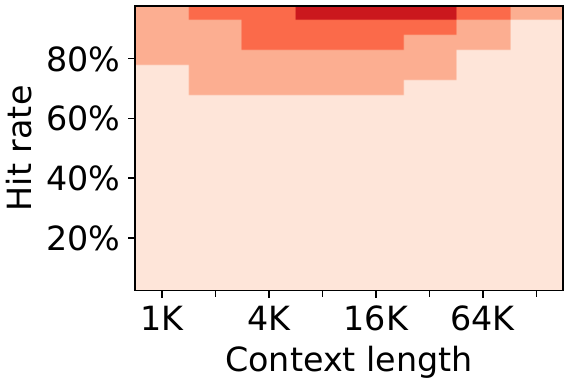}
      \caption{H100, DeepSeek-Distill 7B}
    \end{subfigure}\hfill
    \begin{subfigure}[t]{0.235\textwidth}
      \centering
      \includegraphics[width=\textwidth]{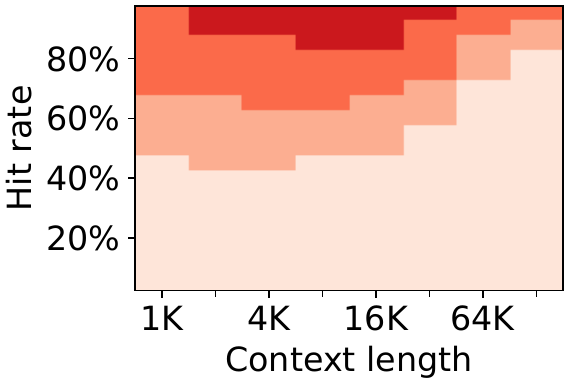}
      \caption{H100, Granite~3.3~8B}
    \end{subfigure}
    \caption{Required per-layer transfer throughput to fully overlap cached KV
    loading with compute across four models on A100 (top row) and H100 (bottom
    row). Faster compute on H100 raises the bandwidth needed for perfect
    overlap.}
    \label{fig:multimodel_required_bw}
\end{figure*}

Figures~\ref{fig:multimodel_kvcache_size}--\ref{fig:multimodel_required_bw}
extend the main-paper recompute and required-throughput surfaces across the
four representative 7--8B models and both GPU platforms.  Total cached KV
size is GPU-agnostic (Figure~\ref{fig:multimodel_kvcache_size}); per-layer
compute time (Figure~\ref{fig:multimodel_compute_time}), MFU
(Figure~\ref{fig:multimodel_mfu}), and required bandwidth
(Figure~\ref{fig:multimodel_required_bw}) each stack A100 (top row) and
H100 (bottom row) to expose the GPU-platform effect at constant model
and context.

\section{Optimal Aggregation Size Benchmark}
\label{app:agg-size-helpers22}

Here we test what is the optimal aggregation size for different chunk granularities.

\begin{figure}[t]
    \centering
    \includegraphics[width=\columnwidth]{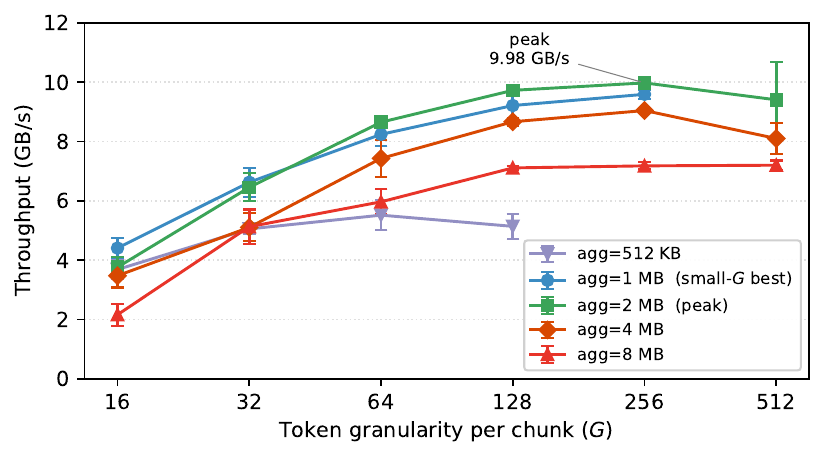}
    \caption{S3RDMA Agg throughput across aggregation sizes and chunk
    granularities. The sweep covers \texttt{agg\_size}~$\in$~\{512KB,1MB,2MB,4MB,8MB\}
    and $G \in \{16, 32, 64, 128, 256, 512\}$. Throughput peaks at
    $G{=}256$ with \texttt{agg\_size}{=}2~MB (9.98~GB/s); for $G{=}16$,
    \texttt{agg\_size}{=}1~MB is best.}
    \label{fig:agg_size_helpers22}
\end{figure}


\FloatBarrier
\section{Reference Algorithms and Cost Table}
\label{app:reference-algs}

This appendix collects the full pseudocode for the ObjectCache layerwise GET and
the Calibrated Stall-opt scheduler, together with the storage-tier cost
breakdown referenced in the main discussion.

\begin{table}[h]
\caption{ObjectCache layerwise GET (full pseudocode).}
\label{alg:layerwise-get}
\centering
\setlength{\tabcolsep}{4pt}
\begin{tabular}{@{}rp{0.83\columnwidth}@{}}
\toprule
1 & \textbf{for} $\ell = 0,\ldots,L-1$ \textbf{do} \\
2 & \quad $B_\ell \leftarrow \emptyset$ \\
3 & \quad \textbf{for each} key $H_j$ in \texttt{chunk\_keys} \textbf{do} \\
4 & \quad\quad $o \leftarrow \ell \cdot S$ \\
5 & \quad\quad append \textsc{RangeGet}$(H_j,o,S)$ to $B_\ell$ \\
6 & \quad \textsc{RDMAWrite}$(\texttt{client\_buffer}[\ell], B_\ell)$ \\
7 & \quad \textsc{NotifyLayerReady}$(\ell)$ \\
\bottomrule
\end{tabular}
\end{table}

\begin{table}[h]
\caption{Calibrated Stall-opt layerwise bandwidth allocation (full pseudocode).}
\label{alg:calibrated-stallopt}
\centering
\setlength{\tabcolsep}{4pt}
\begin{tabular}{@{}rp{0.83\columnwidth}@{}}
\toprule
1 & \textbf{Input:} epoch requests $\mathcal{R}$, bandwidth cap $B$, scheduler margin $\Delta$; per-request bytes-per-layer $b_i$ and per-layer compute time $t_i$. \\
2 & \textbf{for each} request $i \in \mathcal{R}$ \textbf{do} \\
3 & \quad $n_i \leftarrow b_i / t_i + \Delta$ \\
4 & Allocate up to $n_i$ while capacity remains. \\
5 & Redistribute unassigned budget to requests with remaining per-layer stall. \\
6 & Hold per-request rates stable for this epoch. \\
7 & Dispatch layer payloads with weighted deficit round robin. \\
\bottomrule
\end{tabular}
\end{table}

\begin{table}[h]
\caption{Approximate cost comparison across storage tiers (2025 cloud pricing).}
\label{tab:cost}
\centering
\begin{tabular}{lrrr}
\toprule
Tier & Capacity & \$/GB/month & Latency \\
\midrule
GPU VRAM (H100 80GB)   & 80 GB     & \$25+     & $<$1 $\mu$s \\
CPU DRAM               & 512 GB    & \$3--5    & 100 ns \\
Remote DRAM (RDMA)     & TB-scale  & \$1--2    & 10--100 $\mu$s \\
NVMe SSD (local)       & TB-scale  & \$0.10    & 100 $\mu$s \\
S3/Object Storage      & PB-scale  & \$0.02    & 10--50 ms \\
\bottomrule
\end{tabular}
\end{table}

\begin{table}[t]
\centering
\caption{Extra GPU prefill latency per cache-hit boundary when chunk granularity increases from G=16 (vLLM \cite{kwon2023efficient} default) to G=512 (Mooncake \cite{qin2024mooncake}), measured on A100/H100. Each boundary recomputes up to 496 redundant tokens. At 1K QPS, the Llama 8B-64K penalty wastes $\sim$1,500 GPU hours per day for both A100 and H100.}
\label{tab:granularity_comparison}
\renewcommand{\arraystretch}{1.12}
\setlength{\tabcolsep}{3.6pt}
\small
\begin{threeparttable}
\begin{tabular}{@{}lrrrr@{}}
\toprule
& \multicolumn{2}{l}{\textbf{Llama 3.1 8B}} & \multicolumn{2}{l}{\textbf{DeepSeek R1 7B}} \\
\cmidrule(lr){2-3} \cmidrule(lr){4-5}
\textbf{Metric} & \textbf{4K} & \textbf{64K} & \textbf{4K} & \textbf{64K} \\
\midrule
\textbf{\(\Delta\) FLOPs} & 8.0\,T & 24.0\,T & 7.3\,T & 19.5\,T \\
\midrule
\textbf{A100 \(\Delta t\)} & 31.0\(\pm\)5.5\,ms & 63.7\(\pm\)23.7\,ms & 30.3\(\pm\)2.7\,ms & 70.0\(\pm\)6.9\,ms \\
\textbf{H100 \(\Delta t\)} & 18.8\(\pm\)1.0\,ms & 63.4\(\pm\)8.3\,ms & 16.1\(\pm\)1.9\,ms & 40.1\(\pm\)12.4\,ms \\
\bottomrule
\end{tabular}
\begin{tablenotes}[flushleft]
\footnotesize
\item $\Delta$\,FLOPs: extra recompute work per request from the coarser
chunk tail (496 additional tokens), constant across hit rates $r\in\{12.5\%,25\%,\ldots,87.5\%\}$.
Latency rows: mean $\pm$ std of
$\Delta t = T_{G{=}512} - T_{G{=}16}$ measured at each $r$
with one chunk miss. Steady-state after warm-up;
full per-hit breakdowns in Appendix Tables~1--2.
\end{tablenotes}
\end{threeparttable}
\end{table}

\begin{table}[h]
\caption{Client-visible element count for S3RDMA Agg bounded layerwise aggregation.}
\label{tab:lw_opcount}
\centering
\setlength{\tabcolsep}{2.0pt}
\begin{tabular}{@{}lrrrrrr@{}}
\toprule
 & \multicolumn{3}{c}{Context = 4K} & \multicolumn{3}{c}{Context = 64K} \\
\cmidrule(lr){2-4}\cmidrule(lr){5-7}
chunk (tokens)                & 16      & 64    & 256   & 16      & 64      & 256    \\
\midrule
number (hit=87.5\%)          & 224     & 56    & 14    & 3{,}584 & 896     & 224    \\
\addlinespace[2pt]
original elements & 7{,}168 & 1{,}792 & 448 & 114{,}688 & 28{,}672 & 7{,}168 \\
agg payload size              & 1~MB & 2~MB & 2~MB & 1~MB & 2~MB & 2~MB \\
elements per agg              & 16 & 8 & 2 & 16 & 8 & 2 \\
elements after agg            & 448 & 224 & 224 & 7{,}168 & 3{,}584 & 3{,}584 \\
reduction factor              & $16\times$ & $8\times$ & $2\times$ & $16\times$ & $8\times$ & $2\times$ \\
\bottomrule
\end{tabular}
\end{table}

\clearpage
\section{Compute time and required throughput at the canonical configurations}
\label{app:canonical-cells}

Table~\ref{tab:canonical_cells_compute} reports total prefill compute time,
per-layer compute time, and the per-layer transfer throughput needed for
full overlap, for the $(C, r)$ configurations used throughout the
streaming-overlap analysis and scheduler workloads. 

\begin{table}[!h]
\centering
\caption{Per-layer compute time and required transfer throughput to fully
overlap one layer of cached KV loading with compute, for Llama~3.1~8B on
A100~80GB. Total cached KV bytes = $r\cdot C\cdot L\cdot
\mathrm{b}$ with $L{=}32$ layers and $\mathrm{b}{=}4096$ bytes per
token per layer. $T^{(\ell)}_{\mathrm{compute}} =
T_{\mathrm{total}}/L$. Required transfer throughput $= \mathrm{KV}/T_{\mathrm{total}}$
(equivalently per-layer-KV~/~per-layer-compute).}
\label{tab:canonical_cells_compute}
\small
\setlength{\tabcolsep}{3.5pt}
\renewcommand{\arraystretch}{1.05}
\begin{tabular}{lrrrrr}
\toprule
 Context  & Hit  & Cached & $T_{\mathrm{total}}$ & $T^{(\ell)}_{\mathrm{compute}}$ & Req. BW \\
 Length $C$   &  Rate $r$   & Tokens   & (ms)                  & (ms / layer)                    & (GB/s) \\
\midrule
4K   & 0.500 &  2{,}048 &     185.31 &   5.79 & 1.45 \\
4K   & 0.875 &  3{,}584 &      63.47 &   1.98 & 7.41 \\
16K  & 0.500 &  8{,}192 &     955.89 &  29.87 & 1.12 \\
16K  & 0.875 & 14{,}336 &     281.76 &   8.80 & 6.67 \\
32K  & 0.500 & 16{,}384 &   2{,}589.25 &  80.91 & 0.83 \\
32K  & 0.875 & 28{,}672 &     763.19 &  23.85 & 4.92 \\
64K  & 0.500 & 32{,}768 &  8{,}672.79 & 271.02 & 0.50 \\
64K  & 0.875 & 57{,}344 &  2{,}423.90 &  75.75 & 3.10 \\
\bottomrule
\end{tabular}
\vspace{0.25em}

\end{table}

\FloatBarrier

\clearpage
\section{Per-request allocations behind Figure~\ref{fig:lwcpu_scheduler_methods}}
\label{app:per-method-allocations}


Table~\ref{tab:per_method_allocations_16k64k} reports the per-request transfer
throughput each scheduling policy assigns under shared 80~Gbps and 50~Gbps caps
to show moderate and stronger contention for the scheduler workloads.


\begin{table}[!h]
\centering
\caption{Per-request allocated transfer throughput (Gbps) for the
parallel scheduler workloads, using an 80~Gbps cap for Workload~A and
a 50~Gbps cap for Workloads~B and~C.}
\label{tab:per_method_allocations_16k64k}
\small
\setlength{\tabcolsep}{3pt}
\renewcommand{\arraystretch}{1.05}
\begin{tabular}{@{}lrrrrr@{}}
\toprule
Request & Equal & KV-prop & BW-prop & Stall-opt & Cal.\,Stall-opt \\
\midrule
\multicolumn{6}{l}{\textit{Workload A: 16K/64K, 50\%/87.5\%, 80~Gbps cap}} \\
16K, 50\%    & 20.00 &  5.82 &  7.89 &  8.99 & 13.99 \\
16K, 87.5\%  & 20.00 & 10.18 & 46.85 & 42.25 & 27.25 \\
64K, 50\%    & 20.00 & 23.27 &  3.48 &  3.96 &  8.96 \\
64K, 87.5\%  & 20.00 & 40.73 & 21.78 & 24.81 & 29.81 \\
\midrule
\multicolumn{6}{l}{\textit{Workload B: 16K/64K, 50\%/87.5\%, 50~Gbps cap}} \\
16K, 50\%    & 12.50 &  3.64 &  4.93 &  8.99 &  8.26 \\
16K, 87.5\%  & 12.50 &  6.36 & 29.28 & 12.35 & 10.93 \\
64K, 50\%    & 12.50 & 14.55 &  2.17 &  3.96 &  8.96 \\
64K, 87.5\%  & 12.50 & 25.45 & 13.61 & 24.70 & 21.85 \\
\midrule
\multicolumn{6}{l}{\textit{Workload C: 16K/32K/64K, 50\%/87.5\%, 50~Gbps cap}} \\
16K, 50\%    &  8.33 &  2.60 &  3.28 &  5.76 &  4.97 \\
16K, 87.5\%  &  8.33 &  4.55 & 19.45 &  7.62 &  6.58 \\
32K, 50\%    &  8.33 &  5.19 &  2.42 &  6.64 &  7.03 \\
32K, 87.5\%  &  8.33 &  9.09 & 14.36 & 10.78 &  9.30 \\
64K, 50\%    &  8.33 & 10.39 &  1.44 &  3.96 &  8.96 \\
64K, 87.5\%  &  8.33 & 18.18 &  9.04 & 15.24 & 13.15 \\
\bottomrule
\end{tabular}
\end{table}

Table~\ref{tab:s3agg_lw_cap_exact_cells} reports the measured S3Agg-LW-cap
TTFT for the exact allocation configurations in Table~\ref{tab:per_method_allocations_16k64k}.
Table~\ref{tab:s3agg_lw_cap_total_ttft} sums those per-request measurements
into the total TTFT used by the three scheduling workloads.

\begin{table*}[t]
\centering
\caption{Measured S3Agg-LW-cap TTFT for the exact allocation configurations in
Table~\ref{tab:per_method_allocations_16k64k}. }
\label{tab:s3agg_lw_cap_exact_cells}
\small
\setlength{\tabcolsep}{3.5pt}
\renewcommand{\arraystretch}{1.02}
\begin{tabular}{llrrrr}
\toprule
Workload & Policy & Request & Rate & Avg. TTFT & Stdev \\
         &        &         & (Gbps) & (ms) & (ms) \\
\midrule
\multicolumn{6}{l}{\textit{Workload A: 16K + 64K at 50\% and 87.5\%, 80~Gbps cap}} \\
A & Equal & 16K, 50\% & 20.00 & 1,002.9 & 3.9 \\
A & Equal & 16K, 87.5\% & 20.00 & 851.0 & 4.3 \\
A & Equal & 64K, 50\% & 20.00 & 8,188.5 & 29.2 \\
A & Equal & 64K, 87.5\% & 20.00 & 3,330.7 & 16.3 \\
A & KV-prop & 16K, 50\% & 5.82 & 1,592.2 & 1.1 \\
A & KV-prop & 16K, 87.5\% & 10.18 & 1,577.9 & 6.4 \\
A & KV-prop & 64K, 50\% & 23.27 & 8,104.5 & 25.7 \\
A & KV-prop & 64K, 87.5\% & 40.73 & 2,616.0 & 34.1 \\
A & BW-prop & 16K, 50\% & 7.89 & 1,213.8 & 12.9 \\
A & BW-prop & 16K, 87.5\% & 46.85 & 416.7 & 0.9 \\
A & BW-prop & 64K, 50\% & 3.48 & 10,513.9 & 17.2 \\
A & BW-prop & 64K, 87.5\% & 21.78 & 3,086.3 & 3.8 \\
A & Stall-opt & 16K, 50\% & 8.99 & 1,080.4 & 2.1 \\
A & Stall-opt & 16K, 87.5\% & 42.25 & 453.5 & 2.0 \\
A & Stall-opt & 64K, 50\% & 3.96 & 9,312.4 & 4.3 \\
A & Stall-opt & 64K, 87.5\% & 24.81 & 2,754.5 & 14.3 \\
A & Cal. Stall-opt & 16K, 50\% & 13.99 & 1,001.8 & 4.4 \\
A & Cal. Stall-opt & 16K, 87.5\% & 27.25 & 644.7 & 0.3 \\
A & Cal. Stall-opt & 64K, 50\% & 8.96 & 8,531.7 & 3.5 \\
A & Cal. Stall-opt & 64K, 87.5\% & 29.81 & 2,636.3 & 5.2 \\
\midrule
\multicolumn{6}{l}{\textit{Workload B: 16K + 64K at 50\% and 87.5\%}} \\
B & Equal & 16K, 50\% & 12.50 & 1,021.7 & 5.1 \\
B & Equal & 16K, 87.5\% & 12.50 & 1,308.8 & 5.8 \\
B & Equal & 64K, 50\% & 12.50 & 8,448.4 & 8.1 \\
B & Equal & 64K, 87.5\% & 12.50 & 5,139.7 & 14.3 \\
B & KV-prop & 16K, 50\% & 3.64 & 2,489.4 & 5.2 \\
B & KV-prop & 16K, 87.5\% & 6.36 & 2,467.9 & 1.5 \\
B & KV-prop & 64K, 50\% & 14.55 & 8,541.0 & 189.2 \\
B & KV-prop & 64K, 87.5\% & 25.45 & 2,714.4 & 19.4 \\
B & BW-prop & 16K, 50\% & 4.93 & 1,861.4 & 1.9 \\
B & BW-prop & 16K, 87.5\% & 29.28 & 608.8 & 3.0 \\
B & BW-prop & 64K, 50\% & 2.17 & 16,443.2 & 7.9 \\
B & BW-prop & 64K, 87.5\% & 13.61 & 4,751.4 & 12.2 \\
B & Stall-opt & 16K, 50\% & 8.99 & 1,082.5 & 2.7 \\
B & Stall-opt & 16K, 87.5\% & 12.35 & 1,327.8 & 24.1 \\
B & Stall-opt & 64K, 50\% & 3.96 & 9,319.5 & 30.2 \\
B & Stall-opt & 64K, 87.5\% & 24.70 & 2,788.1 & 31.5 \\
B & Cal. Stall-opt & 16K, 50\% & 8.26 & 1,166.8 & 5.9 \\
B & Cal. Stall-opt & 16K, 87.5\% & 10.93 & 1,473.3 & 3.2 \\
B & Cal. Stall-opt & 64K, 50\% & 8.96 & 8,524.8 & 10.8 \\
B & Cal. Stall-opt & 64K, 87.5\% & 21.85 & 3,090.4 & 24.8 \\
\bottomrule
\end{tabular}
\end{table*}

\begin{table*}[t]
\centering
\caption{Measured S3Agg-LW-cap TTFT for the exact allocation configurations in
Table~\ref{tab:per_method_allocations_16k64k}.}
\label{tab:s3agg_lw_cap_exact_cells_workload_c}
\small
\setlength{\tabcolsep}{3.5pt}
\renewcommand{\arraystretch}{1.02}
\begin{tabular}{llrrrr}
\toprule
Workload & Policy & Request & Rate & Avg. TTFT & Stdev \\
         &        &         & (Gbps) & (ms) & (ms) \\
\midrule
\multicolumn{6}{l}{\textit{Workload C: 16K + 32K + 64K at 50\% and 87.5\%, 50~Gbps cap}} \\
C & Equal & 16K, 50\% & 8.33 & 1,167.9 & 17.2 \\
C & Equal & 16K, 87.5\% & 8.33 & 1,914.6 & 12.3 \\
C & Equal & 32K, 50\% & 8.33 & 2,626.0 & 24.4 \\
C & Equal & 32K, 87.5\% & 8.33 & 3,785.7 & 23.6 \\
C & Equal & 64K, 50\% & 8.33 & 8,302.2 & 34.6 \\
C & Equal & 64K, 87.5\% & 8.33 & 7,565.0 & 51.8 \\
C & KV-prop & 16K, 50\% & 2.60 & 3,434.8 & 2.5 \\
C & KV-prop & 16K, 87.5\% & 4.55 & 3,406.4 & 0.1 \\
C & KV-prop & 32K, 50\% & 5.19 & 3,563.8 & 6.5 \\
C & KV-prop & 32K, 87.5\% & 9.09 & 3,475.9 & 10.2 \\
C & KV-prop & 64K, 50\% & 10.39 & 8,216.7 & 455.0 \\
C & KV-prop & 64K, 87.5\% & 18.18 & 3,678.9 & 53.5 \\
C & BW-prop & 16K, 50\% & 3.28 & 2,757.7 & 11.8 \\
C & BW-prop & 16K, 87.5\% & 19.45 & 878.9 & 0.9 \\
C & BW-prop & 32K, 50\% & 2.42 & 7,357.9 & 5.1 \\
C & BW-prop & 32K, 87.5\% & 14.36 & 2,274.2 & 20.8 \\
C & BW-prop & 64K, 50\% & 1.44 & 24,447.9 & 54.1 \\
C & BW-prop & 64K, 87.5\% & 9.04 & 6,981.1 & 25.6 \\
C & Stall-opt & 16K, 50\% & 5.76 & 1,609.5 & 3.5 \\
C & Stall-opt & 16K, 87.5\% & 7.62 & 2,067.5 & 4.1 \\
C & Stall-opt & 32K, 50\% & 6.64 & 2,841.8 & 12.6 \\
C & Stall-opt & 32K, 87.5\% & 10.78 & 2,949.3 & 9.7 \\
C & Stall-opt & 64K, 50\% & 3.96 & 9,301.6 & 28.3 \\
C & Stall-opt & 64K, 87.5\% & 15.24 & 4,299.3 & 52.1 \\
C & Cal. Stall-opt & 16K, 50\% & 4.97 & 1,850.0 & 4.9 \\
C & Cal. Stall-opt & 16K, 87.5\% & 6.58 & 2,381.4 & 2.0 \\
C & Cal. Stall-opt & 32K, 50\% & 7.03 & 2,746.2 & 1.2 \\
C & Cal. Stall-opt & 32K, 87.5\% & 9.30 & 3,403.5 & 25.8 \\
C & Cal. Stall-opt & 64K, 50\% & 8.96 & 8,231.0 & 468.1 \\
C & Cal. Stall-opt & 64K, 87.5\% & 13.15 & 4,892.7 & 1.5 \\
\bottomrule
\end{tabular}
\end{table*}

\begin{table*}[t]
\centering
\caption{Total TTFT for the scheduling workloads using the measured S3Agg-LW-cap
configurations in Tables~\ref{tab:s3agg_lw_cap_exact_cells} and
\ref{tab:s3agg_lw_cap_exact_cells_workload_c}. The baseline is the sum of the
same request configurations without an effective rate limit.}
\label{tab:s3agg_lw_cap_total_ttft}
\small
\setlength{\tabcolsep}{6pt}
\renewcommand{\arraystretch}{1.05}
\begin{tabular}{llrrr}
\toprule
Workload & Policy & Total TTFT & No-limit base & $\Delta$TTFT \\
         &        & (ms)       & (ms)         & (ms) \\
\midrule
A & Equal & 13,373.0 & 12,084.2 & +1,288.8 \\
A & KV-prop & 13,890.5 & 12,084.2 & +1,806.3 \\
A & BW-prop & 15,230.7 & 12,084.2 & +3,146.5 \\
A & Stall-opt & 13,600.9 & 12,084.2 & +1,516.7 \\
A & Cal. Stall-opt & 12,814.5 & 12,084.2 & +730.3 \\
\midrule
B & Equal & 15,918.6 & 12,084.2 & +3,834.4 \\
B & KV-prop & 16,212.6 & 12,084.2 & +4,128.5 \\
B & BW-prop & 23,664.8 & 12,084.2 & +11,580.6 \\
B & Stall-opt & 14,517.9 & 12,084.2 & +2,433.7 \\
B & Cal. Stall-opt & 14,255.3 & 12,084.2 & +2,171.1 \\
\midrule
C & Equal & 25,361.3 & 15,616.1 & +9,745.3 \\
C & KV-prop & 25,776.4 & 15,616.1 & +10,160.3 \\
C & BW-prop & 44,697.7 & 15,616.1 & +29,081.6 \\
C & Stall-opt & 23,069.1 & 15,616.1 & +7,453.0 \\
C & Cal. Stall-opt & 23,505.0 & 15,616.1 & +7,888.9 \\
\bottomrule
\end{tabular}
\end{table*}